\documentclass[runningheads]{llncs}
\usepackage{graphicx}
\usepackage{amsmath,amssymb} % define this before the line numbering.
\usepackage{color}
\usepackage{multirow}

\begin{document}
\pagestyle{headings}
\mainmatter

\def\ACCV20SubNumber{962}  % Insert your submission number here

%===========================================================
\title{AFN: Attentional Feedback Network based 3D Terrain Super-Resolution} % Replace with your title
\titlerunning{AFN}

\author{Ashish Kubade \and
Diptiben Patel \and
Avinash Sharma \and
K. S. Rajan}
\authorrunning{Kubade A, et al.}
\institute{International Institute of Information Technology, Hyderabad, India.
\email{\{ashish.kubade, dipti.patel\}@research.iiit.ac.in},\\
\email{\{asharma, rajan\}@iiit.ac.in}}

\maketitle

%===========================================================
\begin{abstract}
Terrain, representing features of an earth surface, plays a crucial role in many applications such as simulations, route planning, analysis of surface dynamics, computer graphics-based games, entertainment, films, to name a few. With recent advancements in digital technology, these applications demand the presence of high resolution details in the terrain. In this paper, we propose a novel fully convolutional neural network based super-resolution architecture to increase the resolution of low-resolution Digital Elevation Model (LRDEM) with the help of information extracted from the corresponding aerial image as a complementary modality. We perform the super-resolution of LRDEM using an attention based feedback mechanism named `Attentional Feedback Network' (AFN), which selectively fuses the information from LRDEM and aerial image to enhance and infuse the high-frequency features and to produce the terrain realistically . We compare the proposed architecture with existing state-of-the-art DEM super-resolution methods and show that the proposed architecture outperforms enhancing the resolution of input LRDEM accurately and in a realistic manner.
\footnote{Accepted at ACCV'2020.}\footnote{Code for this work is available at \url{https://github.com/ashj9/AFN}}
\end{abstract}

%===========================================================
\section{Introduction}
Real-world terrain is a complex structure consisting of bare land, high range mountains, river paths, arcs, canyons and many more. The terrains and their surface geology are digitally represented using Digital Elevation Models (DEM) or volumetric models. The terrain data coupled with Geographical Information Systems (GIS) extract topological information for various applications including modeling water flow or mass movements, analyse the dynamic behaviour of the earth surface, perform disaster mitigation planning such as flood modeling, landslides, etc. Real-time simulations of terrains are used for fast adaptation and route planning of aerial vehicles such as drones, aircrafts and helicopters, to name a few. Realistic terrain rendering also finds its application in ranging simulations, entertainment, gaming, and many more. As the visual detail and depth in many of these applications, mentioned above, demand terrain information of high resolution and fidelity, capturing or generating such information, as accurately as possible, is the need of the hour.

Diversity and combinations of the complex topological structures make capture/synthesis and analysis of the terrain a challenging task while taking realism into consideration. For instance, computer games with high realistic graphic environments include terrain features for users to experience better realism and allow for detailed exploration. The synthetic or amplified terrain can be used as a background for science fantasy films as well, as the synthetic terrain does not exist and amplified terrain may be difficult for the filming process. 

However, DEMs captured with recent remote sensing sensors are still of relatively low-resolution ($>2$ meters per pixel) and very few geographical locations are captured in high-resolution using airborne LiDAR technology due to high processing requirements. An alternate solution to this problem is to transform the captured low-resolution DEMs (LRDEM) to super-resolved DEMs termed as terrain modeling in general. Existing terrain modeling process can be broadly classified as terrain amplification and terrain synthesis. Terrain amplification enhances the high frequency 3D texture details of the scanned low resolution terrain captured from the real world, thereby making it as close as possible to actual ground truth terrain. On the other hand, terrain synthesis deals with generation of terrain with specific user controls giving a near-realistic appearance. 

\begin{figure}[h]
\centering
\vspace{-0.25in}
\begin{minipage}{0.5\columnwidth}
    \centering
    \includegraphics[width=0.95\columnwidth]{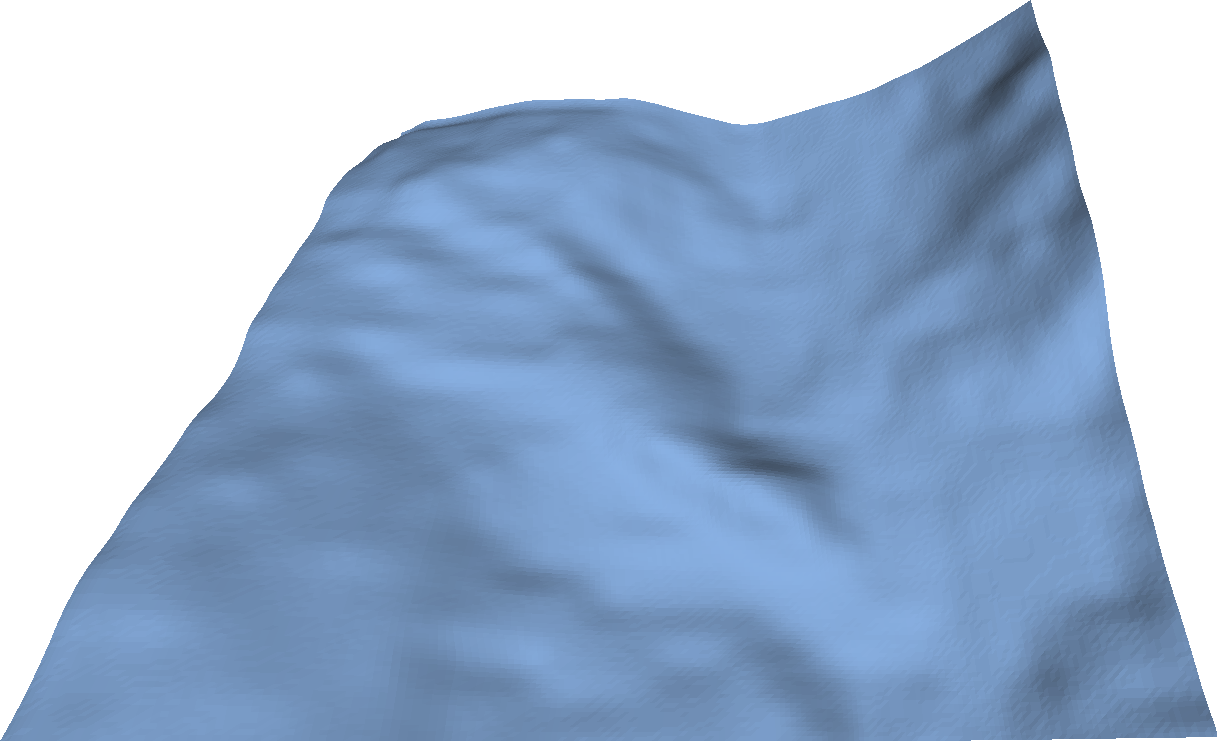} \\ \text{(a) Input LRDEM}
\end{minipage}% 
\begin{minipage}{0.5\columnwidth}
    \centering
    \includegraphics[width=0.95\columnwidth]{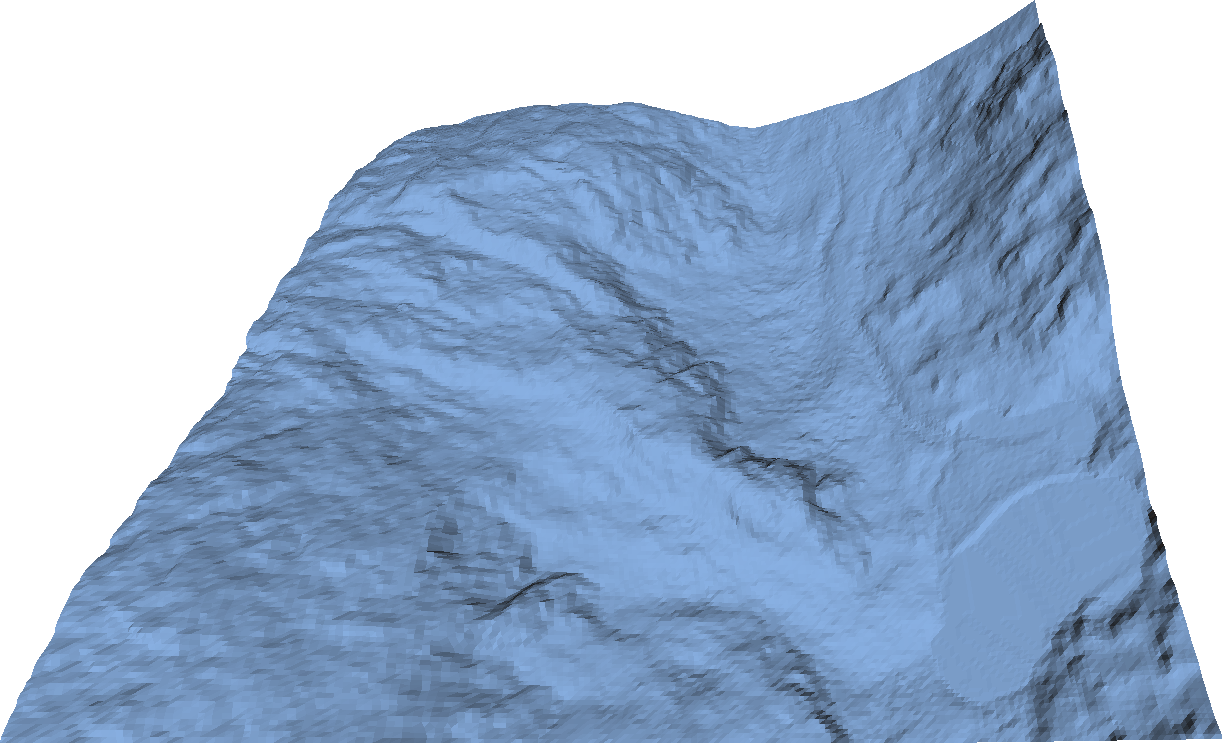} \\ \text{(b) Ground truth HRDEM}
\end{minipage}% 

\begin{minipage}{0.5\columnwidth}
    \centering
    \includegraphics[height=4.5cm,width=0.85\columnwidth]{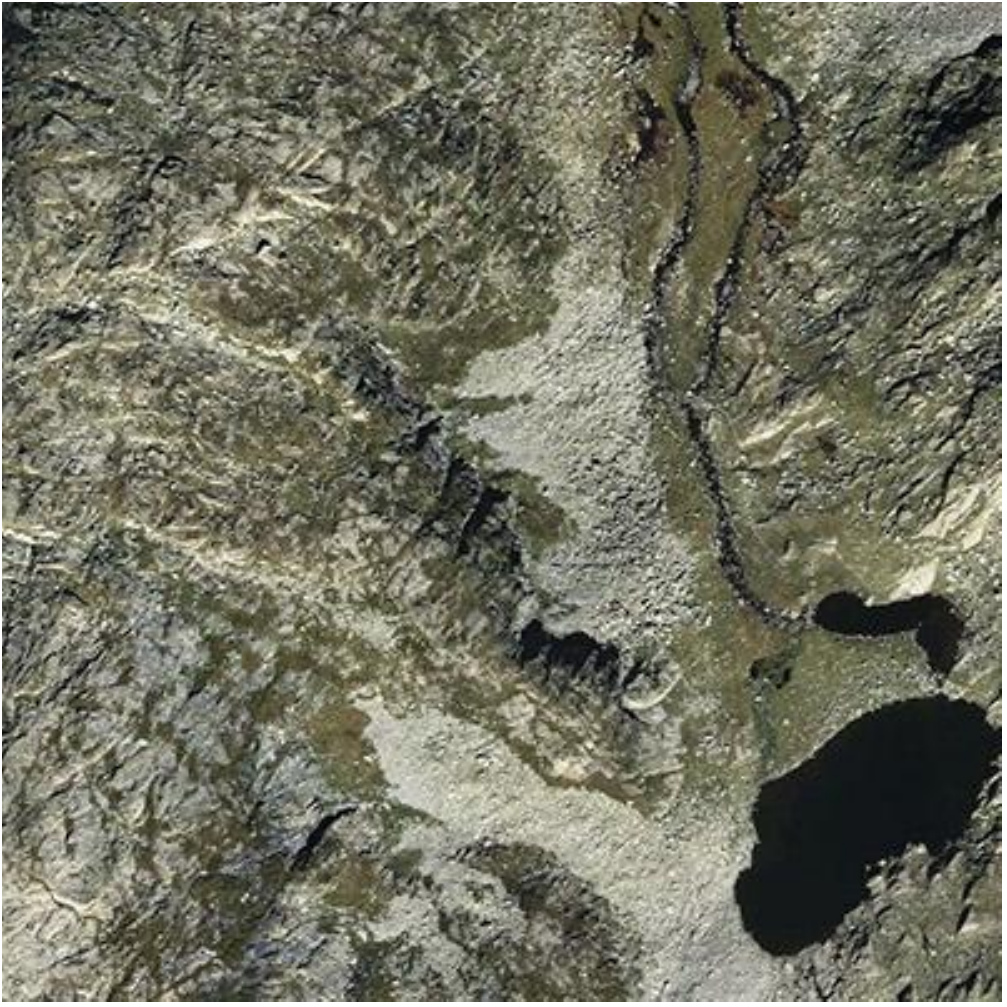} \\ \text{(c) Geo-registered aerial image}
\end{minipage}% 
\begin{minipage}{0.5\columnwidth}
    \centering
    \includegraphics[width=0.95\columnwidth]{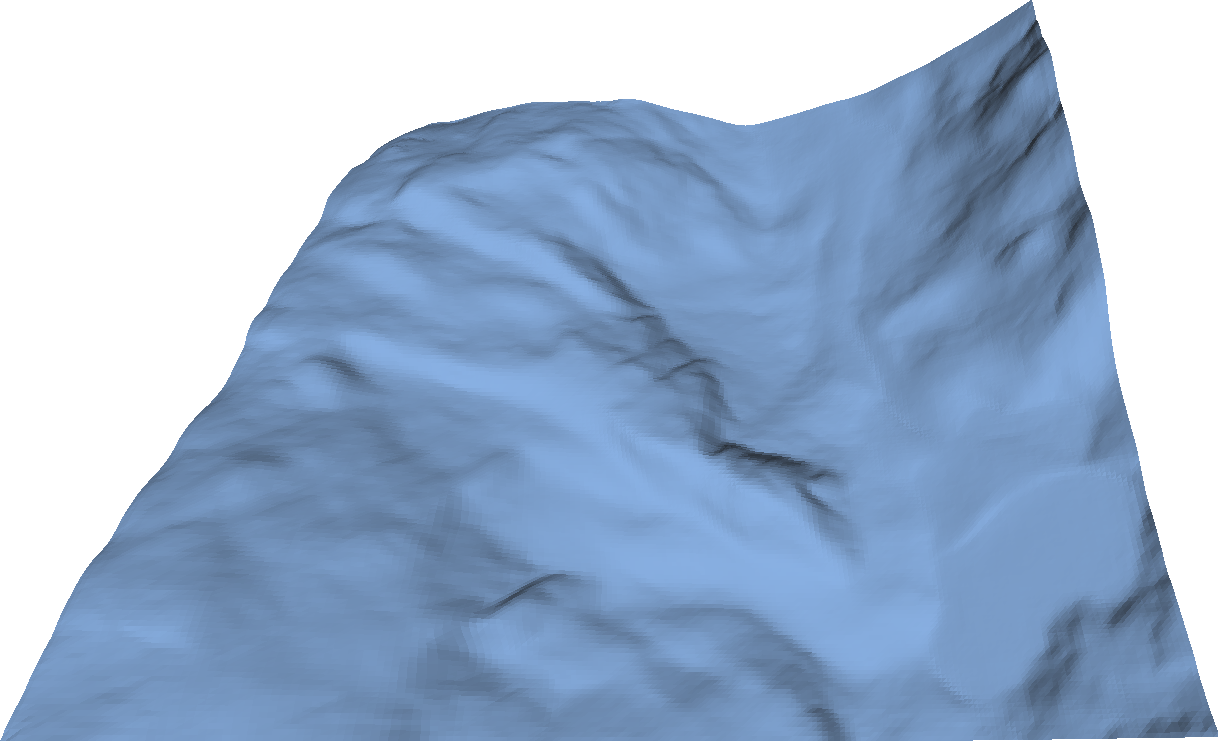} \\ \text{(d) Super-resolved DEM }
\end{minipage}% 
\caption{Views of the terrain at different resolutions and corresponding aerial image.}
\label{fig:inputoutputfig}
\vspace{-0.15in}
\end{figure}

Our primary focus in this work is on terrain amplification of LRDEM (Fig. \ref{fig:inputoutputfig}(a)) with aim to obtain super-resolved DEM (Fig. \ref{fig:inputoutputfig}(d)) terrain models with high fidelity to the ground-truth (Fig. \ref{fig:inputoutputfig}(b)) terrain structures. 

Some of the earliest methods for terrain amplification employed dictionary of exemplars to synthesis high resolution terrains ~\cite{guerin2016sparse,kwatra2003graphcut}, while some other efforts in the literature used erosion simulations to mimic the terrain degradation effects~\cite{musgrave1989synthesis,zhou2007terrain}. Owing to recent advancements in deep learning literature for super-resolution of real world RGB images~\cite{dong2016accelerating,kim2016deeply,zhang2018residual,johnson2016perceptual,Wang_2019,li2019feedback}, some recent efforts have adopted these ideas for DEM super-resolution. DEM Super Resolution with Feedback Block (DSRFB)~\cite{kubade2020feedback} is one such method that attempts to incrementally add high frequency terrain details to the LRDEM in high dimensional feature space using deep learning framework. Another line of work attempted to exploit the terrain information from alternate modalities like aerial (RGB) images (Fig. \ref{fig:inputoutputfig}(c)) that are geo-registered with low resolution DEMs by performing fusion in feature space, e.g.,  Fully Convolutional Networks (FCN) proposed in~\cite{argudo2018terrain}. However, despite using RGB information in DEM super-resolution task, such methods perform poorly in cases of land regions covered with dense vegetation or heavy snowfall. On the other hand, by not availing such modalities (like in DSRFB), we may refrain from exploiting the complementary information captured by RGB images primarily for bare terrain.

% \textcolor{red}{However, despite using RGB information in DEM super-resolution task, such methods \cite{argudo2018terrain} perform poorly in cases of land regions covered with dense vegetation or heavy snowfall. On the other hand, by not availing such modalities (like in DSRFB), we may refrain from exploiting the complementary information captured by RGB images primarily for bare terrain. In contrast, trying to generate a HRDEM from only RGB is inherently difficult as the height cues are almost absent in land covered with snowfall, vegetation or even shadows by larger structures. This motivates us to utilize, these complementary modalities (RGB and LRDEM) in a more efficient and effective manner using the concept of selective fusion in feature space.} 
In this paper, we aim to utilize these complementary modalities in a more efficient and effective manner using the concept of selective fusion in feature space. Attention networks, applied to applications like image captioning \cite{xu2015show}, allow such selective fusion in deep learning framework. Therefore, we aim to design an integrated attention module that enables learning of selective information fusion from multiple modalities. In our setup, where we have two modalities viz aerial image and DEM, we use attention mechanism to selectively pick high frequency details from one modality and discard from the other. Our joint attentional module generates attention mask, which serves as a weight factor deciding the contribution of each modality.

\begin{figure}[t]
    \centering
    \includegraphics[width=1\columnwidth]{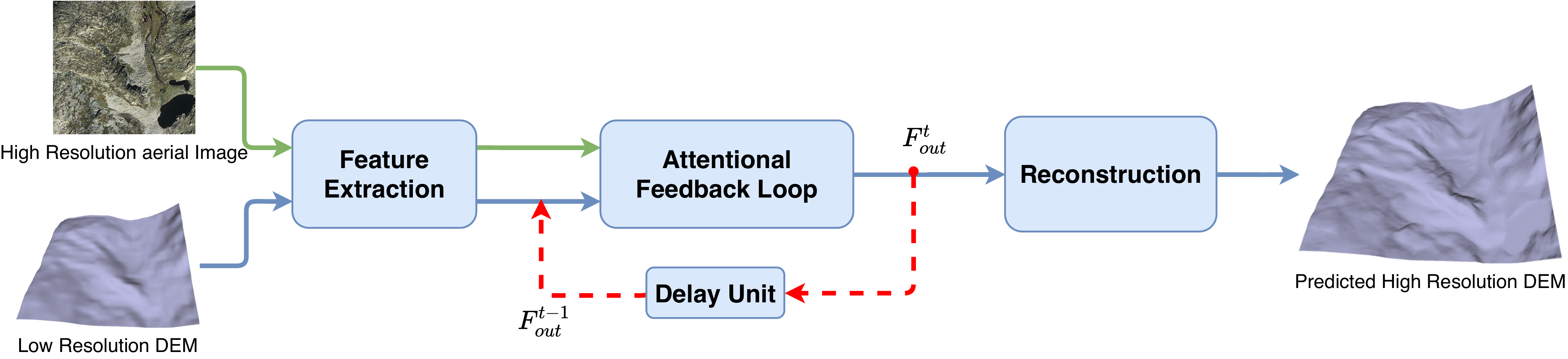}
    \caption{Proposed Attentional Feedback Network Architecture}
    \label{fig:architecture}
    \vspace{-0.15in}
\end{figure}

Thus, we propose a novel terrain amplification method for the DEM representation of real world terrains. We propose supervised learning based fully convolutional neural network (CNN) with LRDEM and corresponding high resolution aerial image as an input and Super-resolved DEM as an output. The architecture of the CNN constitutes a feedback neural network with attention mechanism where the attention mask itself is also allowed to refine its response over the iterations. The high frequency details are added to the DEM using the features extracted from the corresponding high resolution aerial image using the Feature Extraction module. In order to capture high frequency details, we minimize the L1 loss. The overall architecture of the proposed Attentional Feedback Network (AFN) is shown in Fig. \ref{fig:architecture}.

We compare the performance of the proposed methods with other state-of-the-art super-resolution methods for DEM in a quantitative and qualitative manner and are able to achieve better performance in terms of reduced number of parameters as well as inference time. More precisely, proposed AFN solution shares the parameters across feedback loop for incremental fusion in feature space with just 7M parameters whereas other SOTA architectures like \cite{argudo2018terrain} use an order of 20M parameters. Being leaner model, it achieves better performance $50\%$ faster than the average inference time required by \cite{argudo2018terrain} on similar hardware.

%------------------------------------------------------------------------- 
\section{Related Work}
Generating high-resolution of DEM from a low-resolution DEM can be thought of as enhancing or adding the high frequency details like texture patterns, sharp edges, and small curves which often are lost or absent in the low-resolution DEM. With recent success of deep learning, super-resolution of natural RGB images has achieved state-of-the-art performance. However, very few attempts have been made to apply super-resolution to enhance the resolution of DEMs. The possible reasons for fewer attempts could be difference of underlying features, size of features, different textures, and salient objects. Earlier attempts by \cite{argudo2018terrain} have explored the new paths to apply super-resolution to DEMs and successfully demonstrated that deep learning solutions can be adapted to DEMs as well. To understand the challenges in this cross domain task, we would like to highlight some of the major works in respective domains in detail. This section presents a focused overview of terrain modeling, super-resolution methods for images in general and deep learning based feedback network as individual components used in computer vision community. 

\subsection{Terrain Modeling}
Based on the underlying process acquired for terrain modeling, it is classified into three categories: procedural generation methods, physically-based simulation methods, and example-based methods. Procedural generation methods consist of algorithms that use the intrinsic properties of a terrain from the observation of the real world. Physically-based simulation methods execute computer simulations of a geomorphological process that modifies physical properties and surface aspects of a terrain. Example-based methods extract the information from scanned heightfield real world terrains and combine these information for the generation or amplification purpose. The detailed review of existing terrain modeling processes can be referred from \cite{galin2019review}. 

Procedural generation methods use self repeating fractal patterns to mimic the self repeating property of a real world terrain at different scales. Perlin et al. \cite{perlin1985image} proposed the use of generating such fractal patterns for terrain modeling. By using combinations of octaves of noises and thereby creating various scales of noise and smoothness, \cite{musgrave1989synthesis} offers variations in the fractal dimensions. Analogous to mountains, rivers can also be modeled with procedural modeling and incorporated into the landscape \cite{genevaux2013terrain}. User interaction is involved in terrain modeling using painting and brushing on gray-scale images as the fractal’s basis functions for editing in \cite{schneider2006real}. Primitive features in the form of silhouette and shadows, vector based features in the form of ridge lines, riverbeds, cliffs have been used to generate the terrain in \cite{gain2009terrain} and \cite{hnaidi2010feature}, respectively. Hierarchical combination of the primitives such as riverbed, cliffs, hills is used as a tree objects in \cite{genevaux2015terrain}. However, terrains generated using procedural methods lack the effect of natural phenomenon like erosion in their appearances. Hence, a terrain generated by procedural methods is often combined with simulation operations.

Simulation based methods use physical processes such as diffusion, erosion, temperature aided contraction, expansion, hydrological factors aided smoothening, and wind aided gradual abrasion to generate more realistic terrain. \cite{musgrave1989synthesis} presented hydraulic and thermal erosion and combined with ecosystems such as vegetation modeling. However, the heightfield is unable to represent the arches and caves present in the terrain as heightfield can represent only topmost surface in a terrain. \cite{benes2001layered} introduced layered representation for such structures with multiple layers. These structural representations have also enabled stacking multiple layers for effects of various physical and biological phenomenon. One such integration has been represented by \cite{cordonnier2017authoring}, where they fused the interaction between the growing vegetation and terrain erosion by representing them into different layers.

Example-based methods are data-driven methods utilizing the information available in scanned data of real-world terrain. Sample terrain is transformed to desired terrain using user defined sketch in \cite{zhou2007terrain}. Patch based terrain synthesis by using a dictionary of exemplars is performed in \cite{kwatra2003graphcut,guerin2016sparse}. With recently successful deep learning based Generative Adversarial Networks (GANs), \cite{guerin2017interactive} used Conditional GANs to translate a sample terrain using interactive user sketch.
\subsection{Super-resolution of Images}
Different interpolations from neighbourhood information such as linear, bilinear or bicubic are trivial solutions for super-resolution of an image. However, interpolation without high frequency information leads to average out the sharp edges resulting in blur image. Sharp edges and high frequency textures are preserved using Edge Directed Interpolation suggested in \cite{allebach1996edge}. Alternatively, patch based solutions \cite{freeman2000learning,freeman2002example,huang2015single} reconstruct high-resolution patches using a learned mapping between LR and HR patches. While learning the mapping between LR and HR patches, patch consistency is a major issue with patch based approaches. In order to avoid patch inconsistency, mapping between LR and HR images is learned considering an image as a single patch and extracting hand-crafted features using convolutional operators \cite{gu2015convolutional}, gradient profile prior \cite{tai2010super,sun2008image}, Kernel Ridge Regressions (KRR) \cite{kim2010single}.  

Super-resolution task using deep learning is attempted in \cite{wang2015deep,Dong_2016} to learn the mapping between LR and HR. With ResNet overcoming the vanishing gradient problem by using skips connections in deeper networks, super-resolution of images using residual blocks is achieved by DRCN \cite{kim2016deeply}, SRResNet \cite{ledig2017photo}, Residual of Residual (RoR) \cite{zhang2017residual}, Residual Dense Network (RDN) \cite{zhang2018residual}, to name a few. With an emerging interest in generative adversarial networks, super-resolution of an image is attempted by \cite{johnson2016perceptual,zhang2017residual}. While the trend was to go deeper apathetic to the number of parameters, DRRN \cite{tai2017image} formulated a recursive structure to fuse features across all depths.

\subsection{DEM super-resolution with Neural Networks}
Though RDN~\cite{zhang2018residual}, DRRN~\cite{tai2017image} were able to effectively utilize the low level features, the flow of information was only in forward direction, i.e., from initial layers to deeper layers. The low level features are used repeatedly, limiting the reconstruction ability of lower features in the super-resolution task of the network.

SRFBN (Super-Resolution Feedback Network) \cite{li2019feedback} was proposed to tackle this problem. SRFBN used a feedback mechanism adapting from Feedback Networks\cite{zamir2017feedback} in their architecture. Using a feedback mechanism has another advantage with respect to size (number of parameters) of the model. Using a recurrent structure and thereby reusing the parameters has been one of the major techniques in deep learning. Recurrent structures also helps realizing a feedback mechanism easily as recurrent structure can save states of a layer which helps in implementing the feedback component. This approach of super-resolution has been effectively utilized in \cite{kubade2020feedback} for DEM super-resolution task. \cite{kubade2020feedback} have also suggested using overlapped prediction to remove artifacts observed at patch boundaries due to discontinued structures. Even though performing comparable with then state-of-the-art, \cite{kubade2020feedback} network can not avail any additionally available modalities, and hence performance of their method is limited to information cues available in low-resolution DEM only. A Method based on fully convolutional networks by \cite{argudo2018terrain} (referred as FCN, here onwards) extract complementary information from aerial images. However, in their feed-forward setup, there is no control over features learned by initial layers of network. Also, it has been shown that FCN could perform poorly in case of unexposed land regions covered with dense vegetation or areas with heavy snowfall. This motivates us to explore solutions that enable selective extraction of features from aerial images while focusing more on learning of initial layers of the network. We propose the use of attention mechanism for adaptive utilization of features selected respectively from aerial images and DEM. Integrating attention mechanism with feedback network enables the proposed network to learn more refined lower level features.

\section{Method}
Despite using RGB information in super-resolution task, FCN \cite{argudo2018terrain} performs poorly in cases of dense vegetation or heavy snowfall. However, by not availing such modalities, like in DSRFB \cite{kubade2020feedback}, we may refrain ourselves from improvements in super-resolution systems. We utilize these additional modalities in complementary fashion. Inspired from attentional networks applied to applications like image captioning \cite{xu2015show}, we design a module that lets system learn to focus and extract selective information. In our setup, where we have two modalities viz aerial image and DEM, we use attention mechanism to selectively pick high frequency details from one modality and discard from the other. Our joint attentional module generates attention mask, which serves as a weight factor deciding the contribution of each modality.

% Before diving into our network architecture, we briefly discuss the recently proposed feedback based super-resolution architecture termed as SRFBN\cite{li2019feedback} that has enabled to refine the lower level details in the image. 

% We would also like to reiterate, in the proposed method, 
Moreover, our interest is in recovering the lower level details (alternately `high frequency' details) as edges, texture, sharp changes, etc. In a typical Convolutional Neural Network (CNN), these features are captured by the initial layers of the network. To refine the features captured by the shallow layers, we design our attention network in recursive fashion and introduce part of deep features as input back to the shallow layers. This also enables our attention mask to get updated with each time step. Thus, our network becomes a feedback network enabled with attention, we call it as `Attentional Feedback Network' (AFN). The implementation of the feedback module is based on an RNN with $T$ states. With each state, our model refines the lower level features learned by initial layers and enables the reconstruction of SR at each time step. The overall network architecture, once unrolled over time, has the structure as shown in Fig. \ref{fig:unrolled}. In next section, we explain the architectural details of each component.

% However in CNN, information is propagated in only forward direction, i.e. form initial layers (that learn small scale features) towards deeper layers(which learn high level features such as shapes, objects). 

% Feedback networks allow to pass information from deeper layers into initial layers iteratively. For a single example, a feedback network iterates over it for \textit{T} number of steps. In each step, the output of the model is appended to the input of the next step. By introducing the output as feedback from deeper layers to initial layers, we can refine the features learned by initial layers. \cite{dsrfb}DSRFB has effectively adapted this feedback architecture for DEM super-resolution. 

% In proposed method, we integrate attention module into the feedback network. 

\subsection{Proposed Attentional Feedback Network Architecture}
\begin{figure}[t]
\centering
\begin{minipage}{1\columnwidth}
    \centering
    \includegraphics[width=\columnwidth]{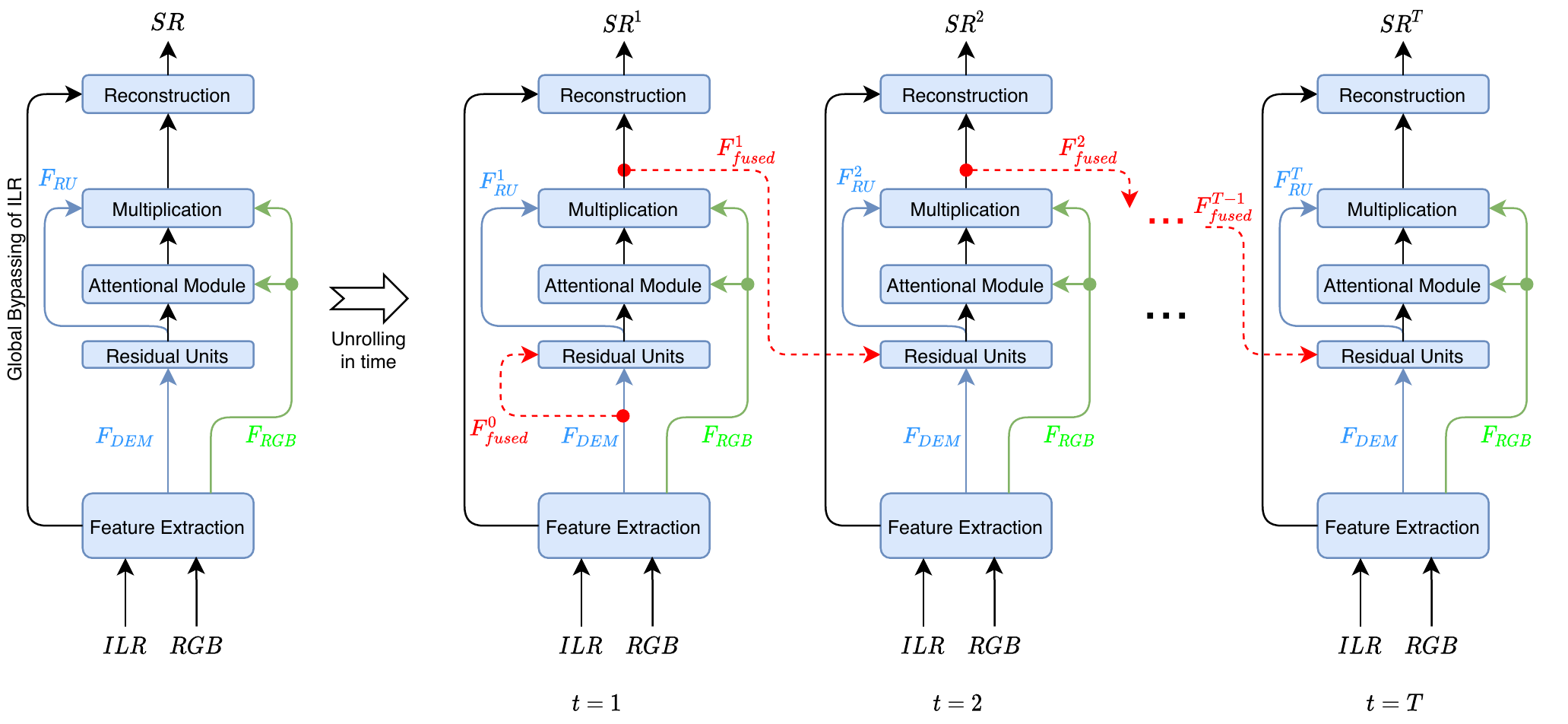}
\end{minipage}
\caption{Unrolled Model Structure}
\label{fig:unrolled}
\vspace{-0.15in}
\end{figure}

As shown in Fig. \ref{fig:unrolled}, unfolded network across time comprises of three components: A Feature Extraction Module, Attentional Feedback Module(AFM) and Reconstruction Block. We also introduce following notations used throughout this paper. 
\begin{itemize}
    \item $m$ denotes the base number of filters
    \item $Conv(m,k)$ denotes a convolutional layer with output number of channels $m$ and kernel size $k$
    \item $T$ denotes the number of steps in feedback loop
\end{itemize}

\begin{figure}[h]
\centering
\vspace{-0.25in}
\begin{minipage}{0.5\columnwidth}
    \centering
    \includegraphics[width=1\columnwidth]{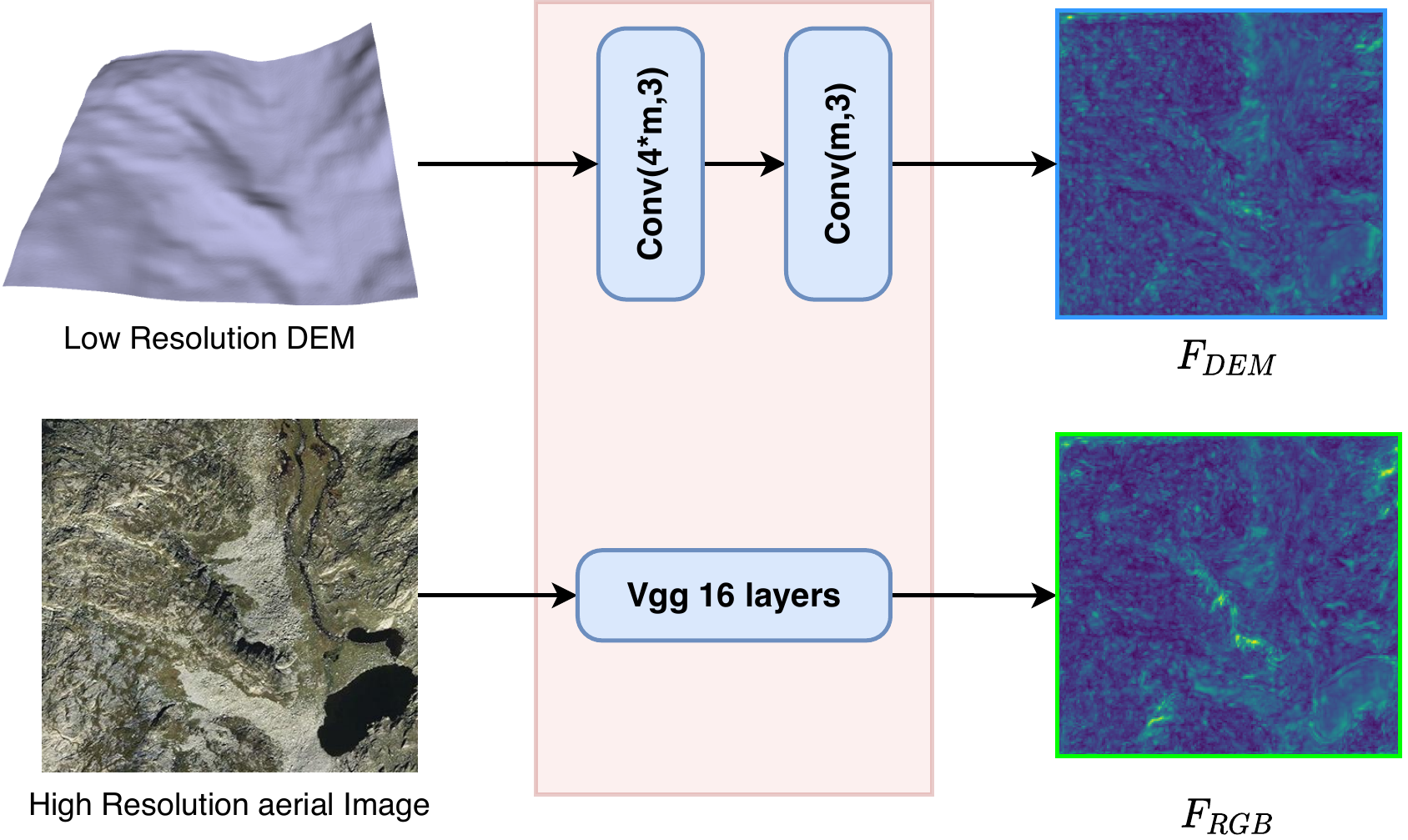}
\end{minipage}%
\caption{Feature Extraction Module}
\label{fig:FE}
\vspace{-0.15in}
\end{figure}

\begin{figure}[h]
\centering
\vspace{-0.25in}
\begin{minipage}{1\columnwidth}
    \centering
    \includegraphics[width=\columnwidth]{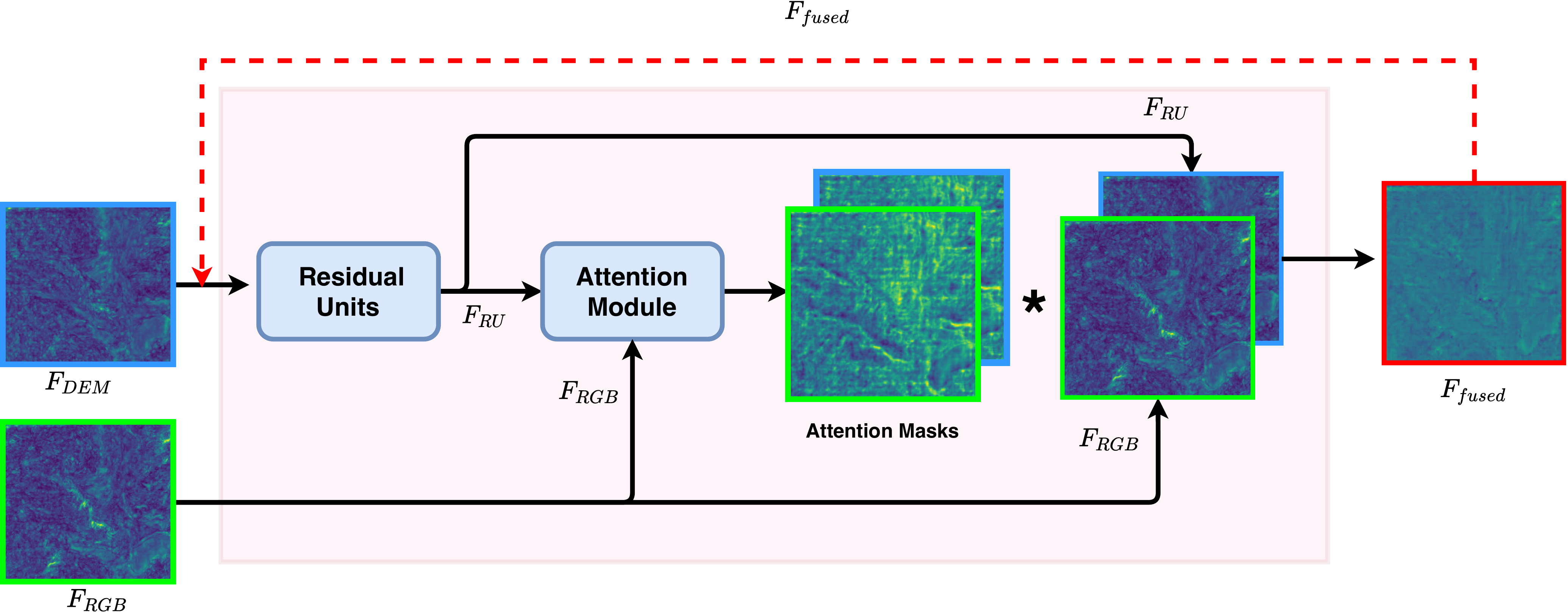} \\ \text{(a) Feedback Module}
\end{minipage}%
\vspace{0.1in}

\begin{minipage}{0.6\columnwidth}
    \centering
    \includegraphics[width=1\columnwidth]{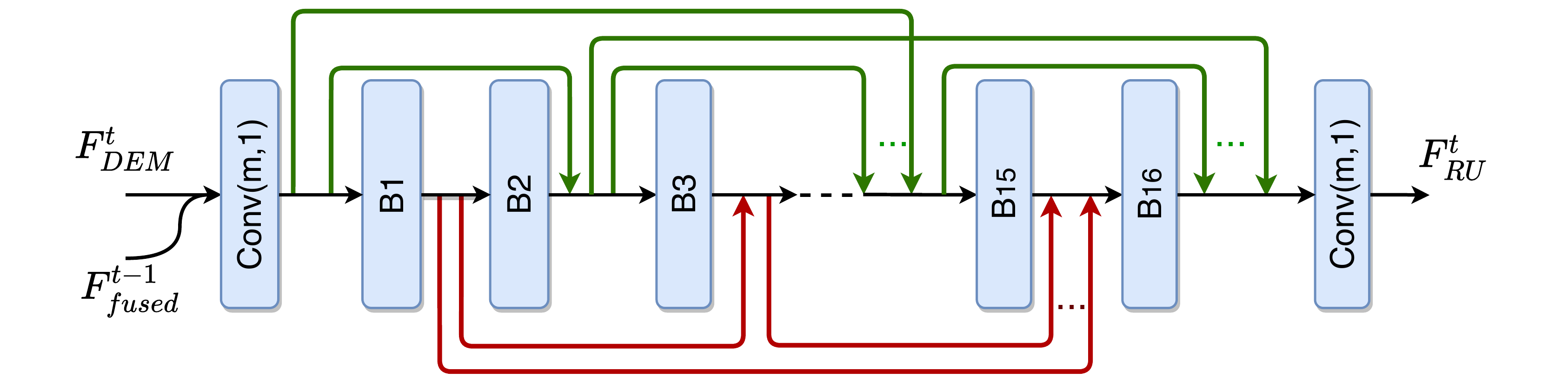} \\ \text{(b) Residual Module}
\end{minipage}%
\begin{minipage}{0.4\columnwidth}
    \centering
    \includegraphics[width=1\columnwidth]{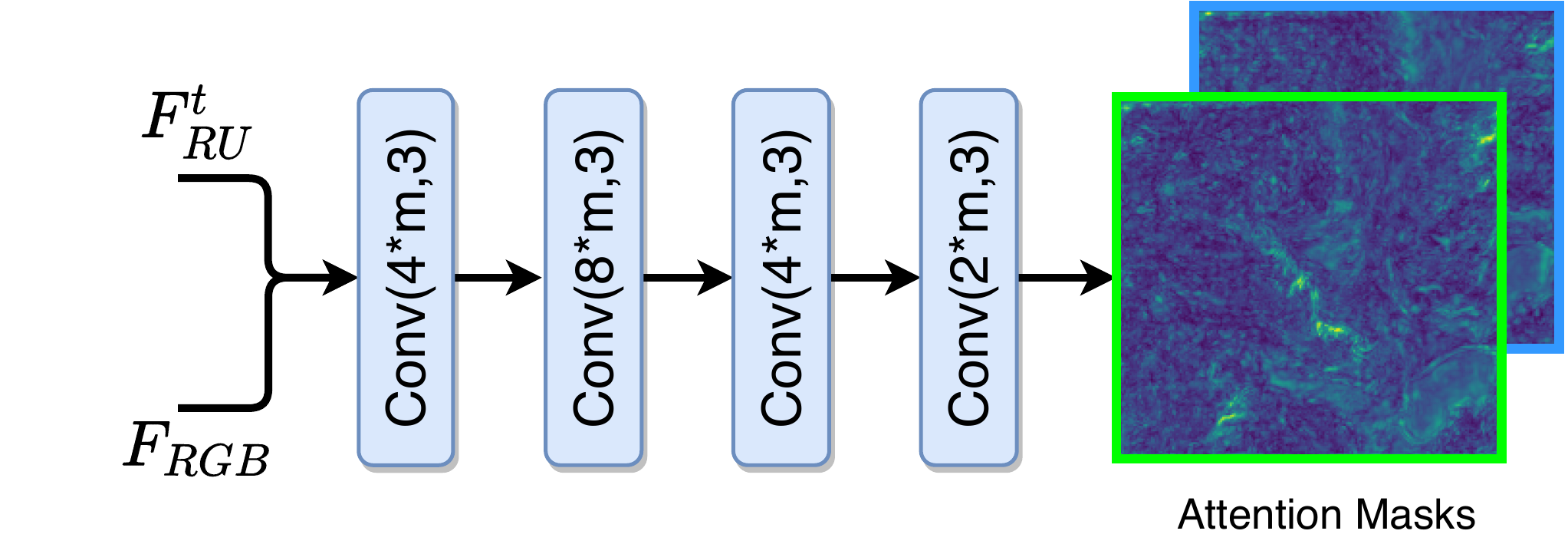} \\ \text{(c) Attention Module}
\end{minipage}%
\caption{Attentional Feedback Module}
\label{fig:AFM}
\vspace{-0.15in}
\end{figure}

 Input to the \textbf{Feature Extraction Module}(FE) is a pair of geo-registered LRDEM and aerial image. As shown in Fig. \ref{fig:FE}, the FE module consists of two branches of layers. Input to the first branch is LRDEM. It comprises of two convolutional layers as $Conv(4*m, 3)$ and $Conv(m, 3)$. The output of this branch is denoted by $F_{DEM}$ (shown with blue outline). Second branch operates on aerial image. We use first two layers from pre-trained VGG-16 network \cite{simonyan2014very} on Imagenet dataset to extract aerial image features. To reduce the domain shift from the aerial images to the images from Imagenet data, we fine-tune these VGG layers during training. The choice of layers has been done empirically by comparing the feature responses of the layers. First two layers are sufficient to extract most of the high frequency details. We denote the output of VGG layers as $F_{RGB}$ (shown with green outline).
 
 We feed the  $F_{DEM}$ and $F_{RGB}$ to \textbf{Attentional Feedback Module}(AFM) which is the heart of our algorithm. As shown in Fig. \ref{fig:AFM}, AFM consists of two sub-modules: A stack of residual units and an attention module. 
 
% We pass the $F_{DEM}$ concatenated with a through the stacked residual units.
Each residual unit consists of a $Conv(m, 1)$ followed by a $Conv(m, 3)$. The $Conv(m,1)$ allows the residual unit to adaptively fuse the information from previous residual units and $Conv(m,3)$ layer produces new $m$ channel features to be passed towards following residual units. 
% of of layers allows to adaptively fues input number of channels to m
The residual units are denoted with $B_{i}$, where $i \in \{1,N\}$, $N$ being an even number. As implemented by \cite{kubade2020feedback}, we use two sets of skip connections to combine the features from residual blocks. The skip connections from $B1$ bypass the information to $\{B2, B4, B6, B8, \ldots, B_{N}\}$, from $B2$ to $\{B3, B5, B7,\ldots, B_{N-1}\}$, from $B3$ to $\{B4, B6, B8, \ldots, B_{N}\}$ and so on. 
% For each data sample, the residual module operates for $T$ number of steps. 
Being inside the iterative feedback module, at each time step $t$, residual module receives a concatenated feature map of $F_{DEM}$ and $F_{fused}^{t-1}$. This timely varying part $F_{fused}^{t-1}$, constitutes the feedback component of our network that we save at time step $t-1$ and is shown as red dashed line in Fig. \ref{fig:AFM}(a).
 A $Conv(m,1)$ layer has been used to compress $F_{DEM}$ and $F_{fused}^{t-1}$ before passing them to B1 at time step $t$. At current iteration, $t$, the outputs from units $\{B2, B4, ..., B_{N} \}$ are compressed by another $Conv(m,1)$ layer to generate the output of residual module viz $F_{RU}^{t}$. 
 
%  Similar to \cite{kubade2020feedback} we use two paths of skip connections as denoted by green and red lines. 

% The $Conv(m,1)$ layers which implicitly implement 1x1 convolution, have been used to adaptively fuse the information from incoming large number of features.

At each time step, $t$, the resultant output of residual module, denoted as $F_{RU}^{t}$, along with the features from the RGB branch i.e. $F_{RGB}$, are fed to the attention module. 
 
 %\par The design of this attention module is inspired from the \cite{li2017instance}. To our task, our attention masks can be thought of as spatial probability maps that we can learn with fully convolutional networks. Thus we design our attentional module mimicking a small fully convolutional network of $4$ layers.
 
Inspired from \cite{li2017instance}, attention masks generated from the attention module can be thought of as spatial probability maps. These spatial probability maps can be learnt using fully convolutional networks. Hence, attention module comprises of a small fully convolutional network of $4$ layers.
 
%  however, we use a set of convolutional layers instead of Matrix multiplication used by \cite{zhang2019self}. 

As shown in Fig. \ref{fig:AFM}(c), the attention module consists of $Conv(4*m, 3), Conv(4*m, 3), Conv(8*m,3)$ and $Conv(2*m, 3)$. The final output with $2*m$ channels has been split into two units: $Attn_{DEM}^{t}$ and $Attn_{RGB}^{t}$ of m channels each which in turn act as an attention mask for the input features $F_{RU}^{t}$ and $F_{RGB}$, respectively. Unlike \cite{li2017instance}, we use multi channel attention maps. We then use element wise channel multiplication to get a  weighted set of features with attention channels. A channel-wise summation then fuses the two sets of features together into, $F_{fused}^{t}$, the final output of AFM as shown in Eq. (\ref{fusion}).
 \begin{equation}
F_{fused}^{t} = F_{RU}^{t} * Attn_{DEM}^{t} + \gamma * F_{RGB} * Attn_{RGB}^{t}
    \label{fusion}
    % check the order once
 \end{equation}
 where a learnable parameter $\gamma$ is used for stable learning. $\gamma$ has been initialized with 0 so as to focus on $F_{RU}$ first and adaptively move the attention to $F_{RGB}$.
  To implement an iterative feedback, we store $F_{fused}^{t}$ over current step and then concatenate it with $F_{DEM}$ to be processed in next step as part of feedback. For the first step, i.e. at $t=0$, as there will not be any $F_{fused}$, we use $F_{DEM}$ itself as feedback information for step $t=0$. We forward $F_{fused}^{t}$ as input to the reconstruction block. Residing inside the feedback module, we let the attention maps to refine themselves as the iterations proceed. This timely varying attention units for same input also makes our attention module unique and different from \cite{li2017instance}.
 
 We run the AFM module for $T$ number of steps. For each step $t$, we get one set of features $F_{fused}^{t}$, which is improved version of itself as the iteration goes on.  
%  For convenience, we denote the fused features at step $t (\in T)$ as $F_{fused}^{t}$.
 
 We implement \textbf{Reconstruction Block} with two units of convolutional layers $Conv(m,3)$ and $Conv(1,3)$. For each step of the feedback unit, the reconstruction block takes in $F_{fused}^{t}$ and produces a residual map denoted by $I_{res}^{t}$. The $I_{res}^{t}$ are the higher frequency details we are interested in generating.
 We add this residual, $I_{res}^{t}$ to $DEM_{ILR}$ which we forward from input directly via a global skip connection shown in Fig. \ref{fig:unrolled}. The predicted super-resolved DEM at time step $t$ is given by Eq. (\ref{eq:SR_t}).
 \begin{equation}
     SR^{t} = I_{res}^{t} + DEM_{ILR} ~~~     \forall   t \in \{1,T\}
     \label{eq:SR_t}
  \end{equation}
 With a recursion of depth $T$, for each step of $t$ for single data instance, we get one SR, forming an array of  predicted SRDEMs with increasing amount of details.
 
 We use $L1$ loss over HRDEM and ${SR}^t$ for $t \in \{1,T\}$ as given by Eq. (\ref{mainLoss}).
 \begin{equation}
     L = \sum_{t=1}^{T} \lvert HRDEM - {SR}^{t} \rvert
     \label{mainLoss}
 \end{equation}
 The final loss $L$ will be used for back-propagation and training the parameters.
%  Given two modalities, attention from one of them can be found by using a convolutional layer. We concatenate the features from 

\section{Experimental Setup}
\subsection{Datasets}
% Superresolution of DEM has been tried by using information from other modalities.

Our goal in this study is to selectively utilize the information from other modalities like aerial images. For fair comparison with existing methods such as \cite{argudo2018terrain} and \cite{kubade2020feedback}, we train our model using dataset provided by Institut Cartogràfic i Geològic de Catalunya (ICC) \cite{icc} and Südtiroler Bürgernetz GeoKatalog (SBG) \cite{sbg}. The terrains provided by these institutes have been pre-processed by the authors of \cite{argudo2018terrain}. The dataset used for training comprises of geo-registered pairs of DEM and aerial images of several mountain regions named Pyrenees and Tyrol. DEM patches with a resolution of 2m/pixel have been used as ground truth (HRDEM) elevation maps. These HRDEMs have been downsampled to 15m/pixel to create a corresponding LRDEM. For convenient training, original DEM tiles have been split into patches of size 200x200 pixels, where each pixel intensity signifies terrain height. To effectively avail the aerial information, the resolution of aerial image has been set twice that of DEM, resulting in patches of size 400x400. From all the patches, 22000 patches have been chosen for training and 11000 patches for validation. Also, two regions from Pyrenees namely Bassiero and Forcanada, and two regions from Tyrol namely Durrenstein and Monte Magro have been set aside for testing the network performance. We suggest the reader to refer \cite{argudo2018terrain} for more details about the dataset.

% \textcolor{red}{To test the robustness of our method, we use one more set of tiles for testing. We use another dataset for testing Cartosat}.

\subsection{Implementation Details}
\begin{figure}[h]
\centering
\vspace{-0.25in}
\begin{minipage}{0.5\columnwidth}
    \centering
    \includegraphics[width=0.95\columnwidth]{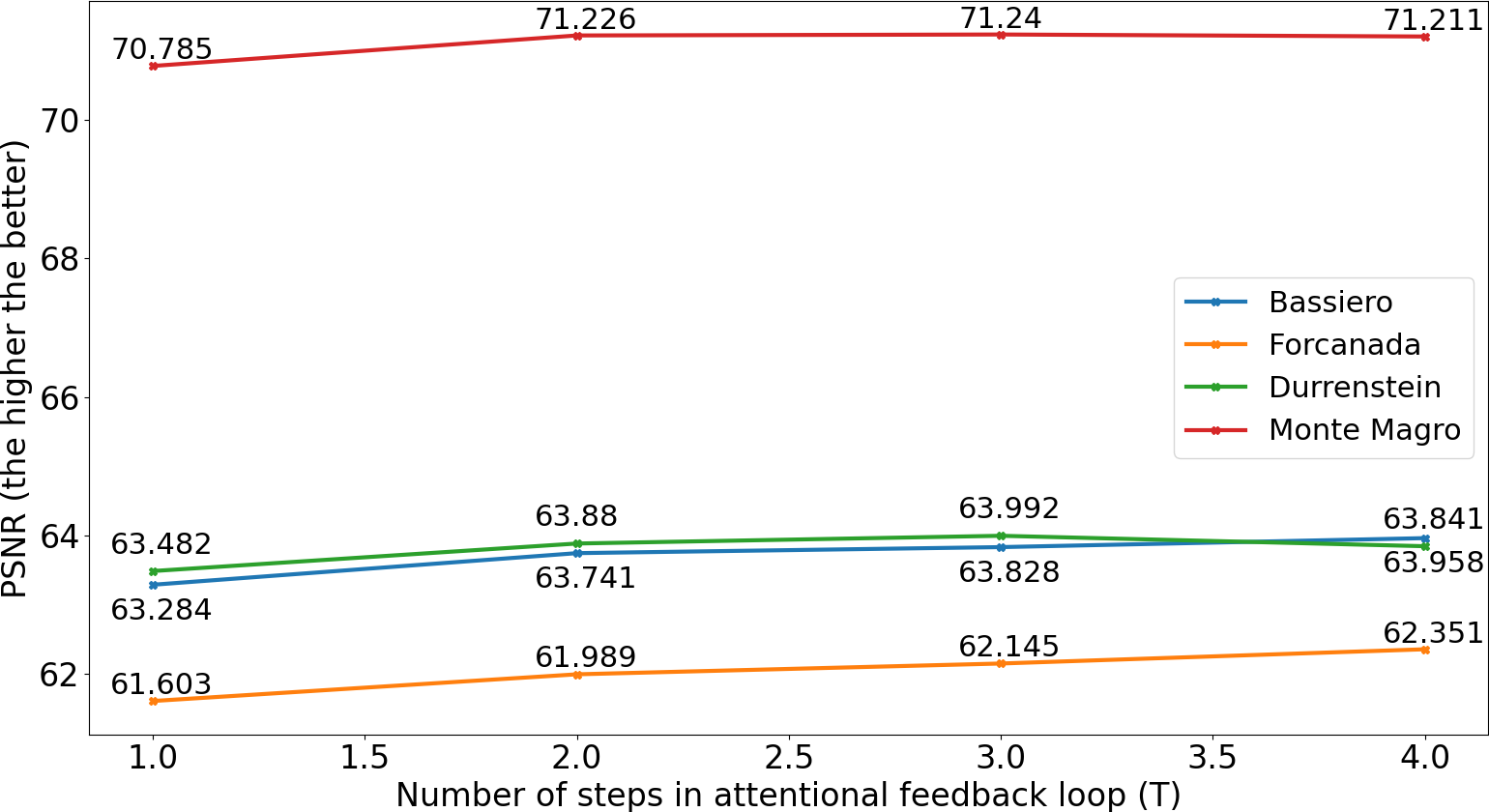} \\ \text{(a) PNSR}
\end{minipage}%
\begin{minipage}{0.5\columnwidth}
    \centering
    \includegraphics[width=0.95\columnwidth]{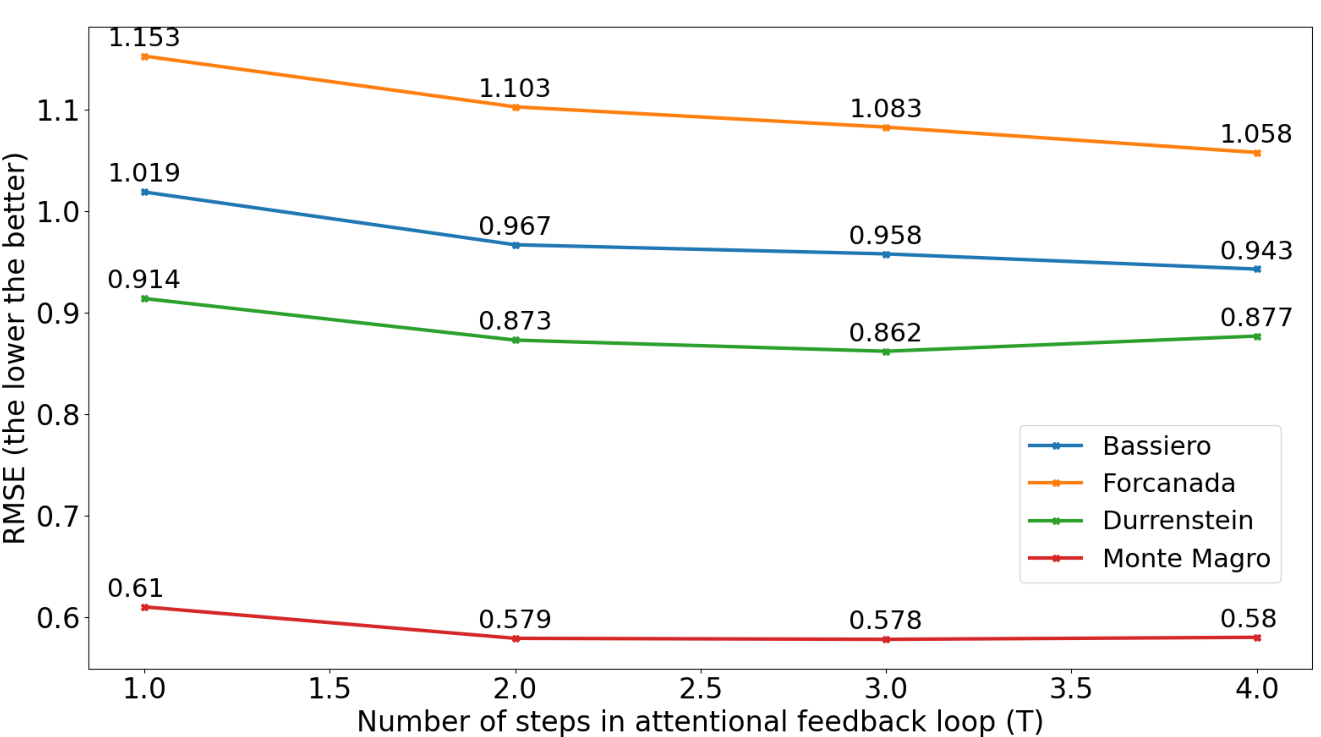} \\ \text{(b) RMSE}
\end{minipage}%
\caption{Choice of parameter $T$ (number of steps)}
\label{fig:T_choice}
\vspace{-0.15in}
\end{figure}

In this section, we explain the hyper-parameters and details about our experimental setup. We have used convolutional layers with kernel size of $3 \times 3$, unless explicitly stated. The parameters in these layers were initialized with \textit{Kaiming} initialization protocol. All the convolutional layers are followed by PReLU activation. For the RGB branch in FE module, we have used first two convolution layers (pre-trained on ImageNet dataset) from VGG-16 network. Later, we allow to fine-tune their weights so as to adapt the weights according to DEM modality.
We set $m$ (the number of base channels) to $64$ and $T$(number of steps in feedback loop) to $4$. We have chosen $T$ to be $4$, as the gain performance in terms of PSNR and RMSE (Shown in Fig. \ref{fig:T_choice}) is getting stagnant around $T = 4$.   
We use $N$, i.e. the number of residual units, as 16. Since we have used LRDEM with resolution of 15 meters (as stated in \cite{kubade2020feedback}), the effective super-resolution factor in our case is 7.5X. We have used a batch size of 4, the max supported with our 4 NVIDIA-1080Ti GPUs.
We used learning rate of $\eta = 0.0001$ with multi-step degradation by parameter $0.5$ with epoch intervals at [45,60,70]. Parameters were updated with $Adam$ optimizer. We have implemented our network in PyTorch framework. After convergence of the network, the value learned by $\gamma$ is 0.358.
% The trained models, code will be made publicly available along with a working demo.

During testing, similar to \cite{kubade2020feedback}, we have adopted the technique of overlapped prediction with overlap of $25\%$ on all sides of the patch.
%%%%%%%%%%%%%%%%%
%%%%%%%%%%%%%%%%%%%%%%%%%%%%
\section{Results and Discussions}
We use standard root mean squared error(RMSE) and peak signal-to-noise ratio (PSNR) metrics to compare the performance of our proposed method with existing SOTA methods, namely FCN\cite{argudo2018terrain} and DSRFB~\cite{kubade2020feedback}. While RMSE helps understand the cumulative squared error between the prediction and ground truth, PSNR helps to gain the measure of peak error, PSNR and RMSE are complementary measures to compare the performance of SR methods. We also compare the performance with a variant of FCN, FCND which does not use aerial imagery as complementary source of information. %without using  DSRFB\cite{kubade2020feedback}.

%\subsection{Accuracy of Generated Terrain}
From Table \ref{table:rmse}, we can infer that our network AFN outperforms both FCN and DSRFB. Using the overlapped prediction, our variant, AFNO performs even better. Similar observation can be made from Table \ref{table:psnr}, where AFN has the best PSNR even without using overlapped prediction. Even though the quantitative performance in some areas seems marginal, the gains achieved by our method (over SOTA) in terms of absolute height values are around 0.5 to 1.0m which is quite valuable.
% Additionally, for terrains with high altitude variations, ... 
\begin{table}[h]
\vspace{-0.2in}
\caption{Comparison: RMSE values(in meters. The lower the better).}
\centering
\resizebox{\textwidth}{!}{
\begin{tabular}{|c@{\hskip2pt}|c|c|c@{\hskip2pt}|c|c|c|c|}
\hline
Input & \multicolumn{4}{c|}{Only LRDEM} & \multicolumn{3}{c|}{LRDEM and RGB} \\
\hline
Region      & Bicubic   & DSRFB & DSRFO & FCND & FCN & AFN & AFNO \\ \hline
Bassiero    &1.406 &	1.146	&1.091	&1.083 & 1.005 & \textbf{0.943} & \textbf{0.926} \\ \hline
Forcanada   & 1.632 & 1.326 & 1.2702 & 1.259 & 1.097 & \textbf{1.058} & \textbf{1.030} \\ \hline
Durrenstein & 1.445 & 0.957 & 0.884 & 0.868  & 0.901 & \textbf{0.877} & \textbf{0.854}\\ \hline
Monte Magro & 0.917 & 0.632 & 0.589 & 0.581  & 0.587 & \textbf{0.580} & \textbf{0.566} \\ \hline
\end{tabular}
}
\label{table:rmse}
\vspace{-0.15in}
\end{table}

\begin{figure}[h]
\centering
%\vspace{-0.25in}
\begin{minipage}{0.095\columnwidth}
    \centering
    \text{ } \\ 
\end{minipage}%
\begin{minipage}{0.23\columnwidth}
    \centering
    \text{(a) Bassiero} 
\end{minipage}%
\begin{minipage}{0.23\columnwidth}
    \centering
    \text{(b) Forcanada}
\end{minipage}%
\begin{minipage}{0.23\columnwidth}
    \centering
    \text{(c) Durrenstein} 
\end{minipage}%
\begin{minipage}{0.23\columnwidth}
    \centering
    \text{(d) Monte Magro}
\end{minipage}%

\begin{minipage}{0.095\columnwidth}
    \centering
    \text{Aerial} \\ 
    \text{Image}
\end{minipage}%
\begin{minipage}{0.23\columnwidth}%for rgb pimages
    \centering
    \includegraphics[width=0.95\columnwidth]{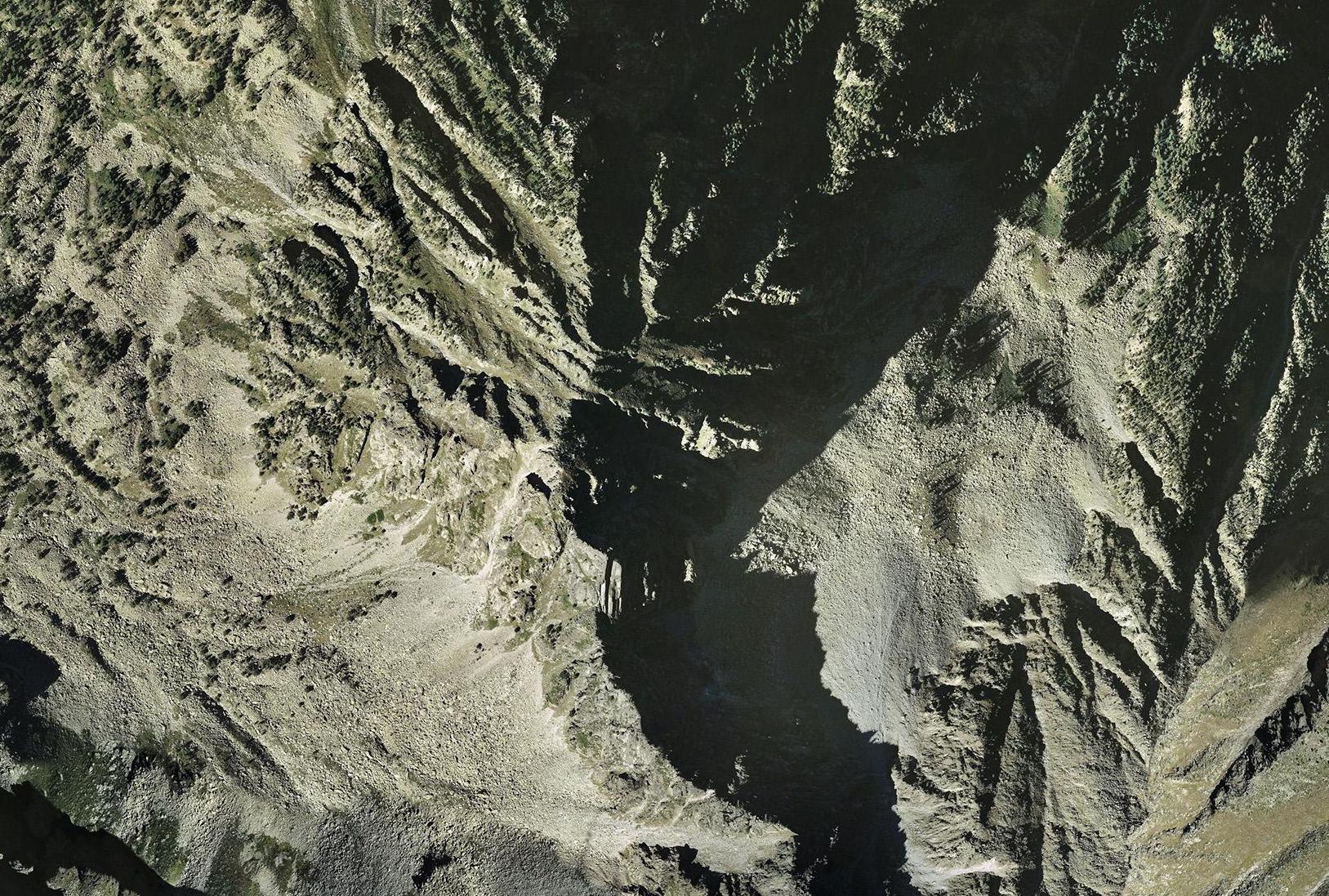} 
\end{minipage}%
\begin{minipage}{0.23\columnwidth}
    \centering
    \includegraphics[width=0.95\columnwidth]{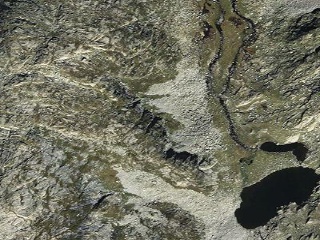}
\end{minipage}%
\begin{minipage}{0.23\columnwidth}
    \centering
    \includegraphics[width=0.95\columnwidth]{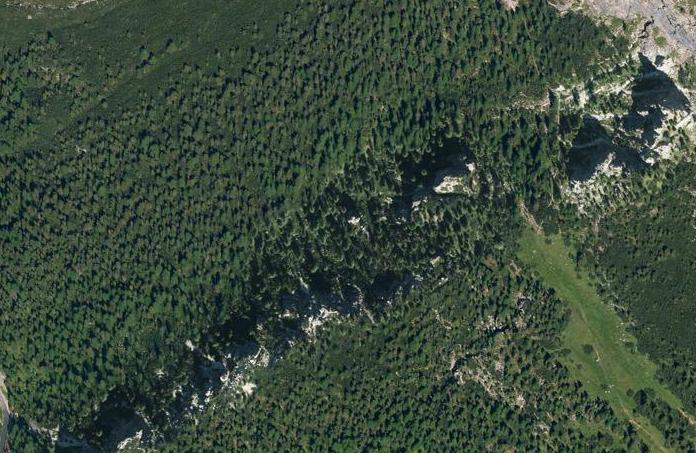}
\end{minipage}%
\begin{minipage}{0.23\columnwidth}
    \centering
    \includegraphics[width=0.95\columnwidth]{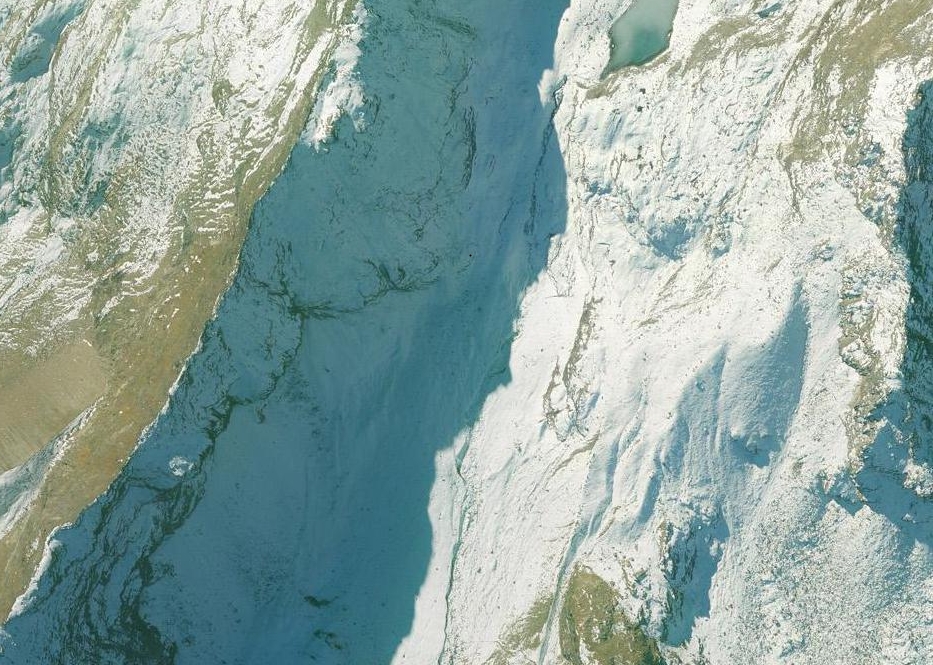}
\end{minipage}% %%%%rgb end
%\centering %%%%%%%LR start

\begin{minipage}{0.095\columnwidth}
    \centering
    \text{LRDEM}
\end{minipage}%
\begin{minipage}{0.26\columnwidth}
    \centering
    \includegraphics[width=1\columnwidth]{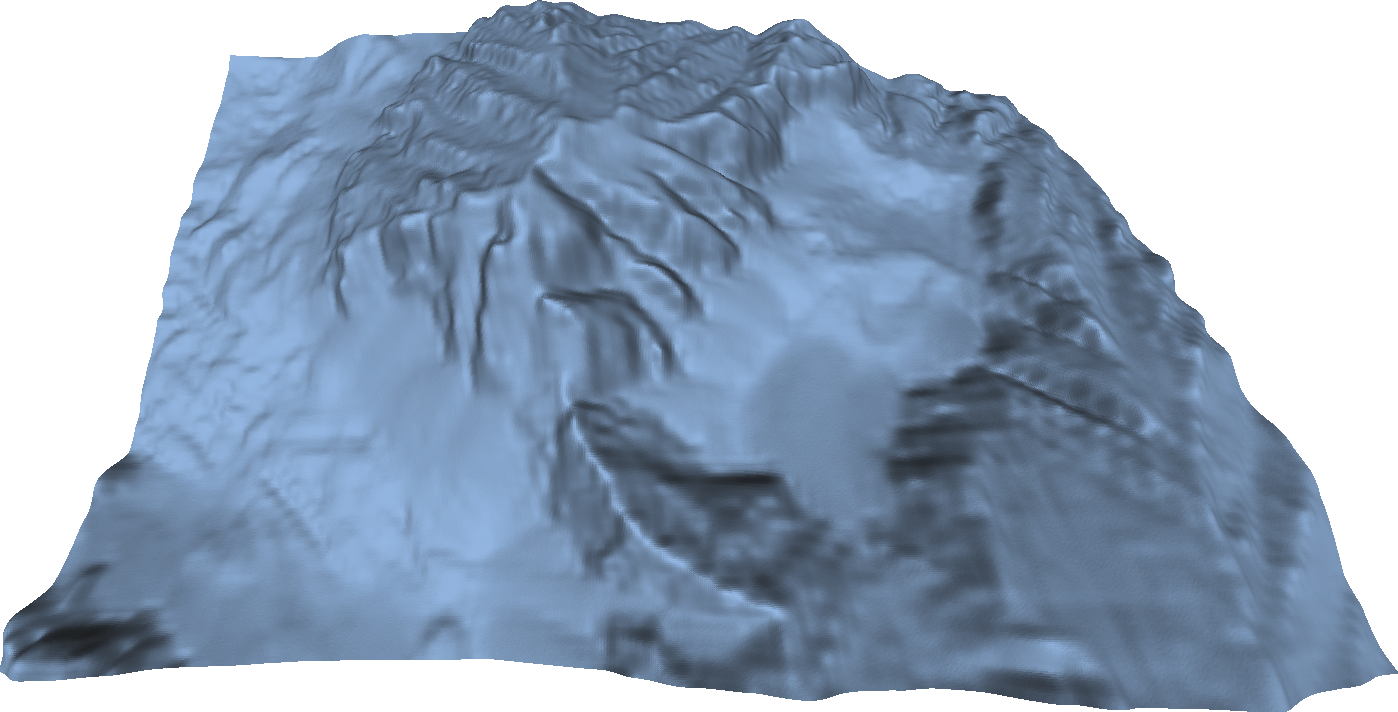}
\end{minipage}%
\begin{minipage}{0.22\columnwidth}
    \centering
    \includegraphics[width=1\columnwidth]{imgs/forc/forc_lr01.png}
\end{minipage}%
\begin{minipage}{0.22\columnwidth}
    \centering
    \includegraphics[width=1\columnwidth]{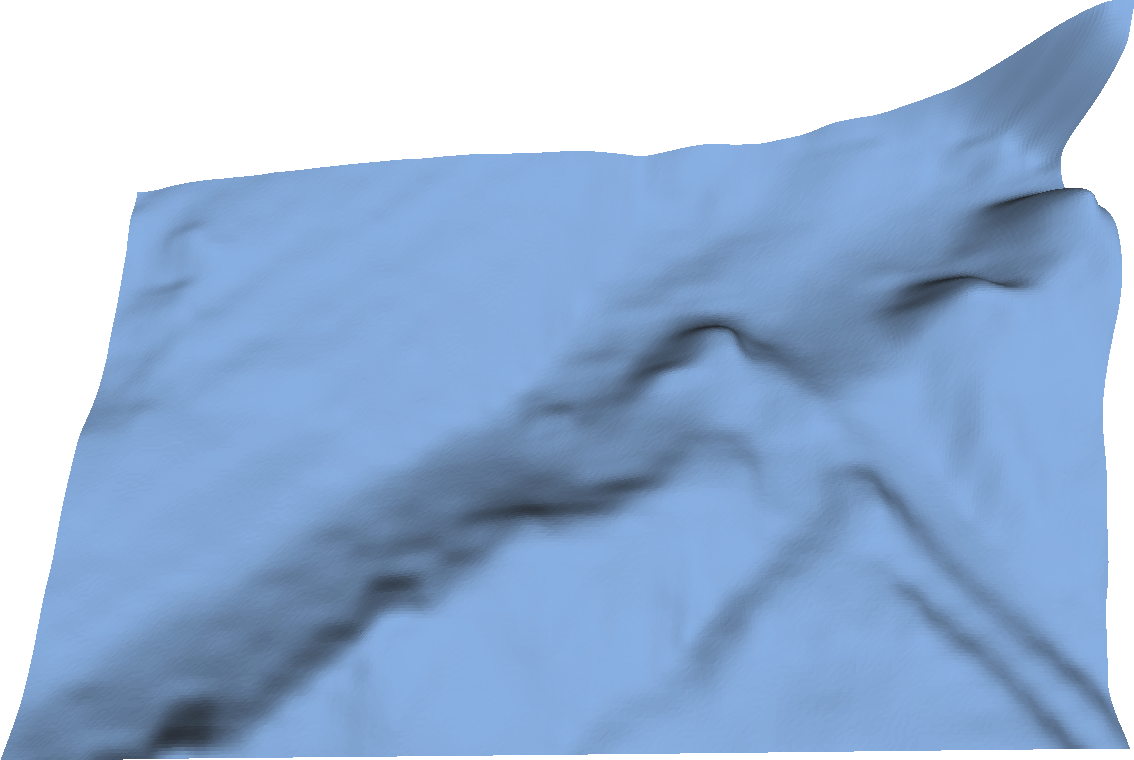}
\end{minipage}%
\begin{minipage}{0.22\columnwidth}
    \centering
    \includegraphics[width=1\columnwidth]{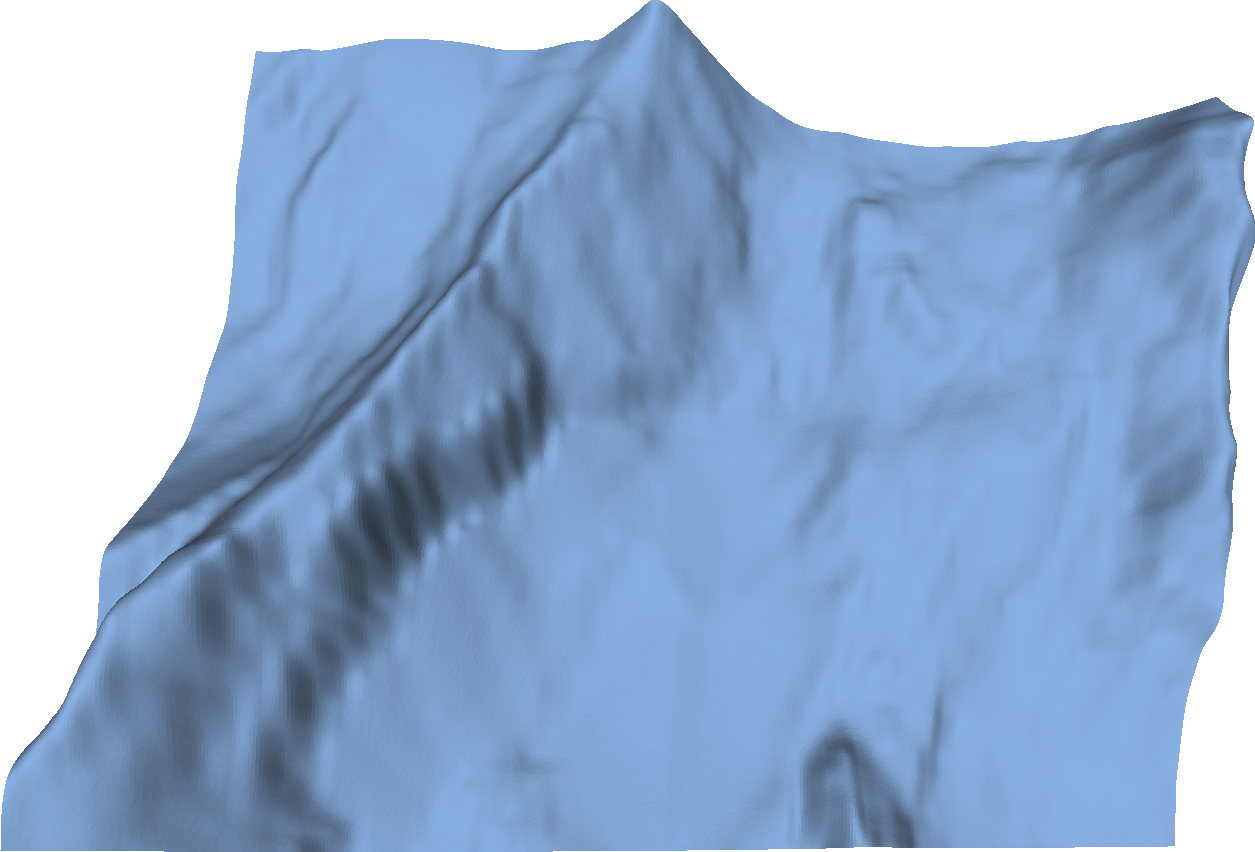}
\end{minipage}% %%%%%%%LR end

%%%%dsrfb start
\begin{minipage}{0.095\columnwidth}
    \centering
    \text{DSRFB}
\end{minipage}%
\begin{minipage}{0.26\columnwidth}
    \centering
    \includegraphics[width=1\columnwidth]{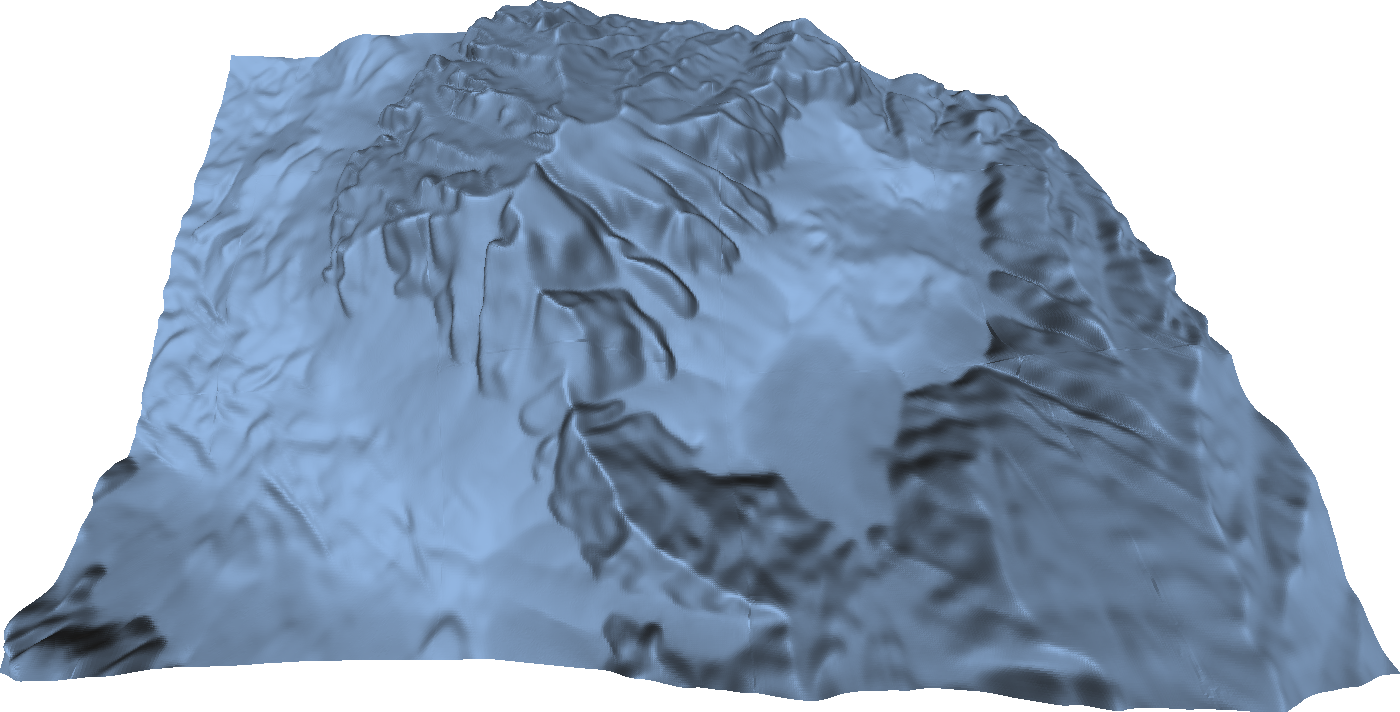}
\end{minipage}%
\begin{minipage}{0.22\columnwidth}
    \centering
    \includegraphics[width=1\columnwidth]{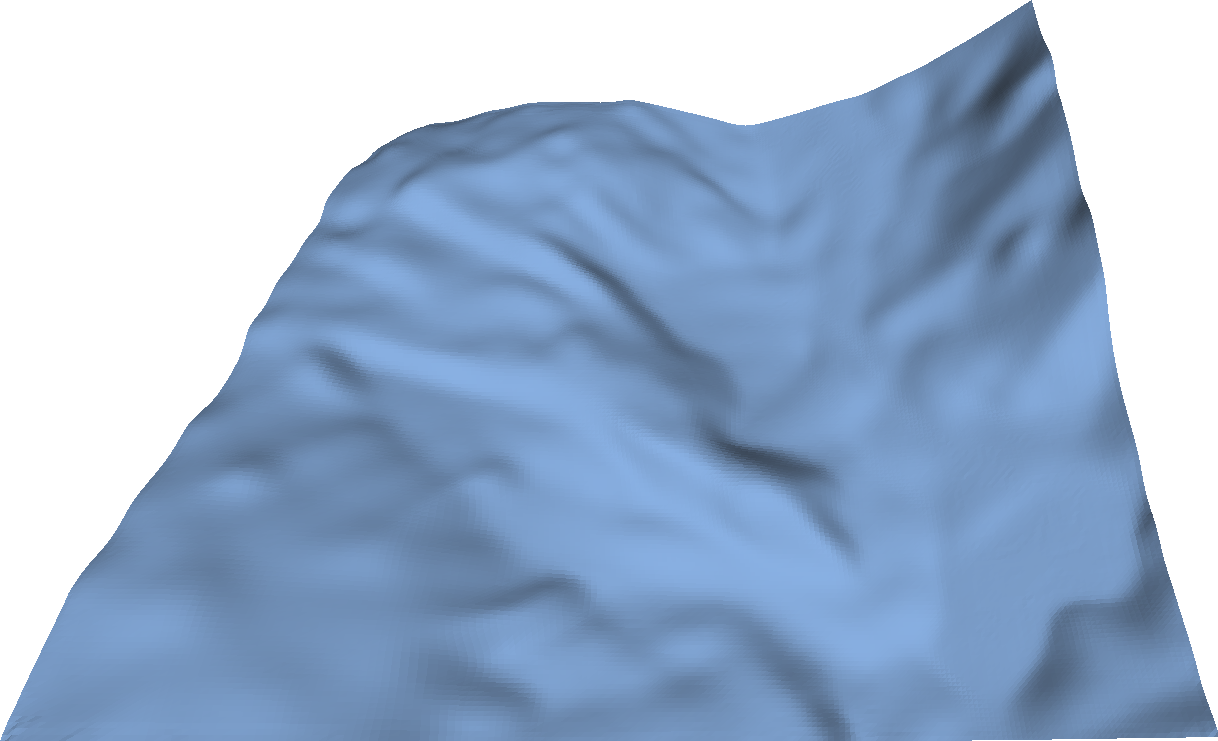}
\end{minipage}%
\begin{minipage}{0.22\columnwidth}
    \centering
    \includegraphics[width=1\columnwidth]{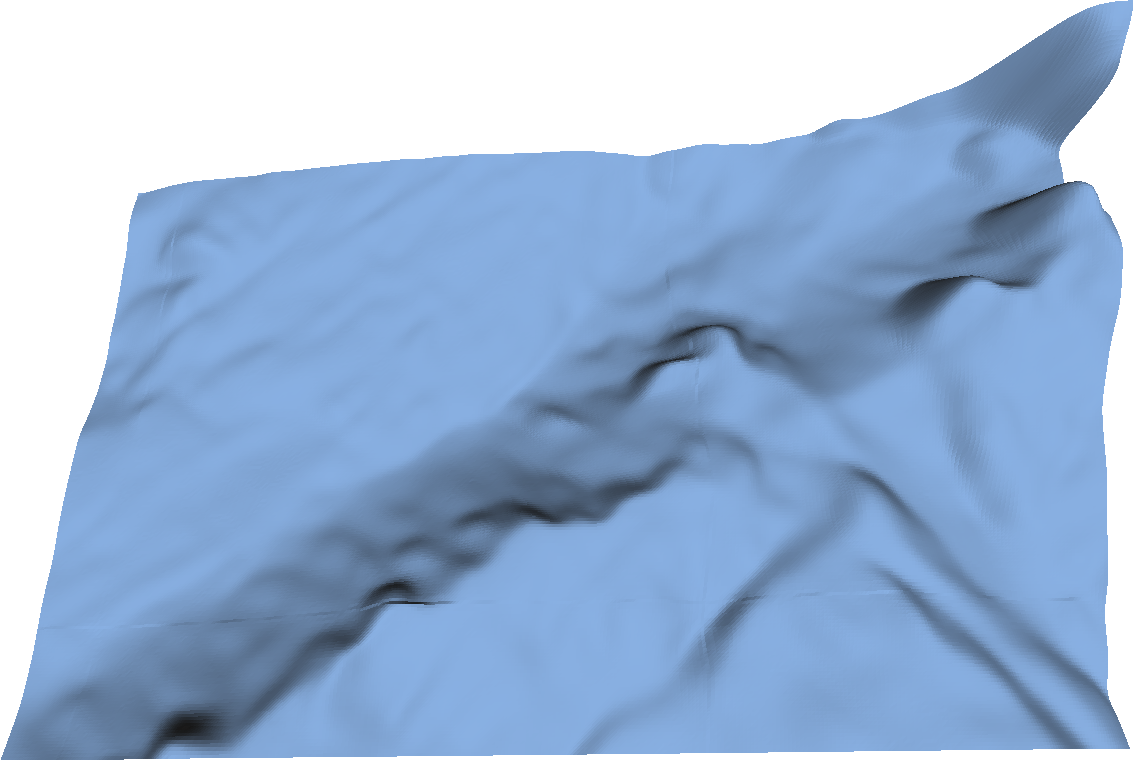}
\end{minipage}%
\begin{minipage}{0.22\columnwidth}
    \centering
    \includegraphics[width=1\columnwidth]{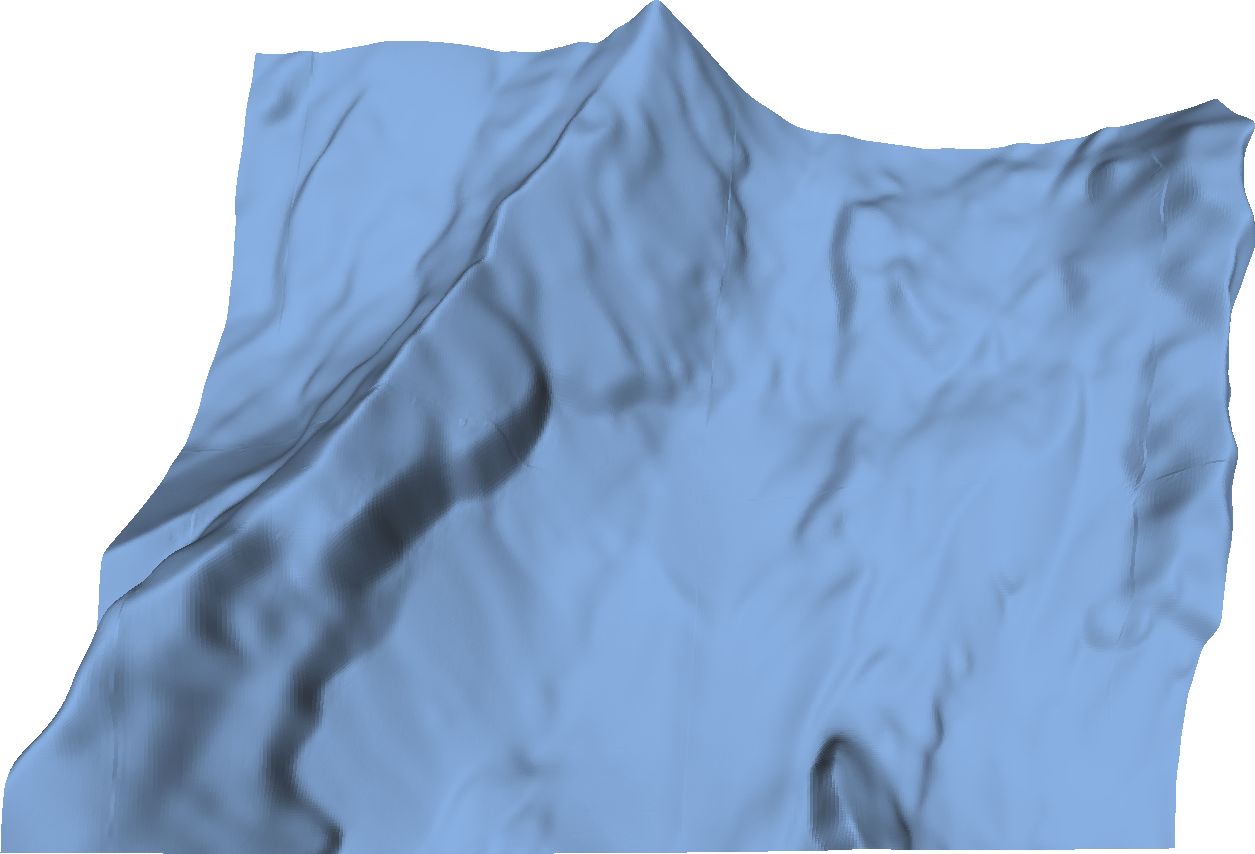}
\end{minipage}%

%for fcn
\begin{minipage}{0.095\columnwidth}
    \centering
    \text{FCN}
\end{minipage}%
\begin{minipage}{0.26\columnwidth}
    \centering
    \includegraphics[width=1\columnwidth]{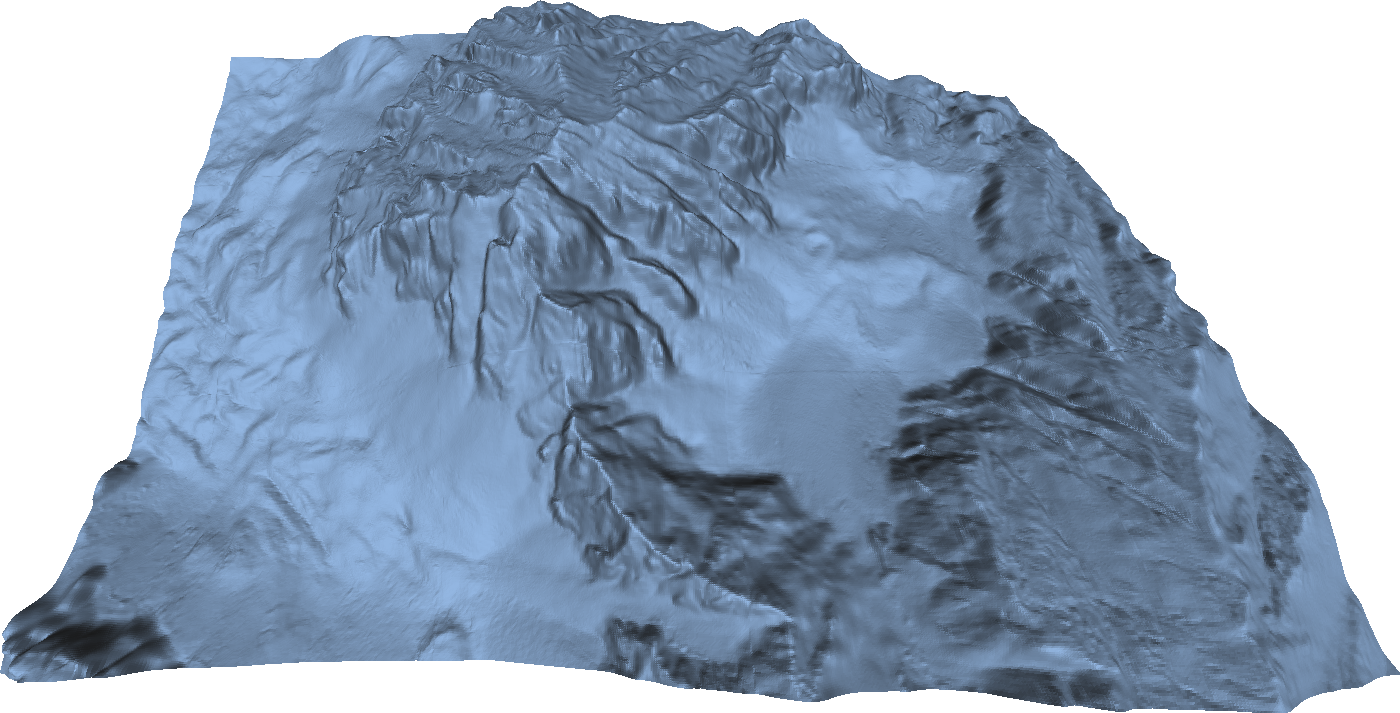}
\end{minipage}%
\begin{minipage}{0.22\columnwidth}
    \centering
    \includegraphics[width=1\columnwidth]{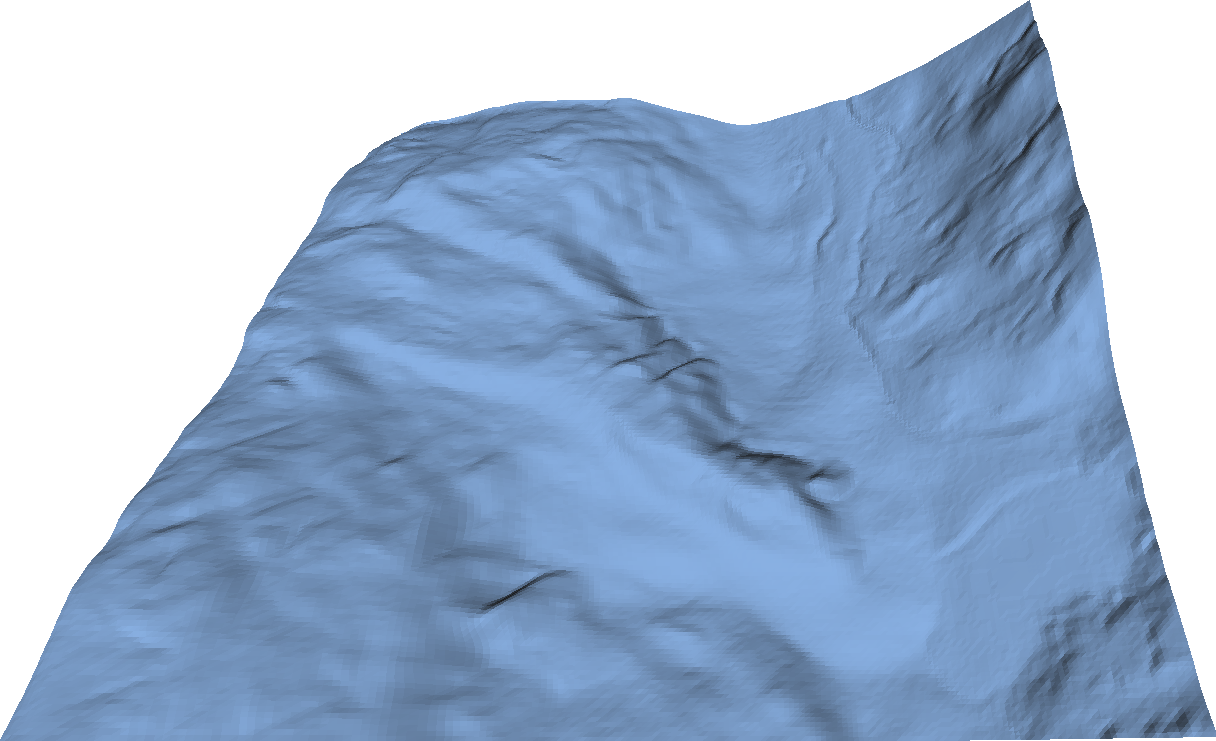}
\end{minipage}%
\begin{minipage}{0.22\columnwidth}
    \centering
    \includegraphics[width=1\columnwidth]{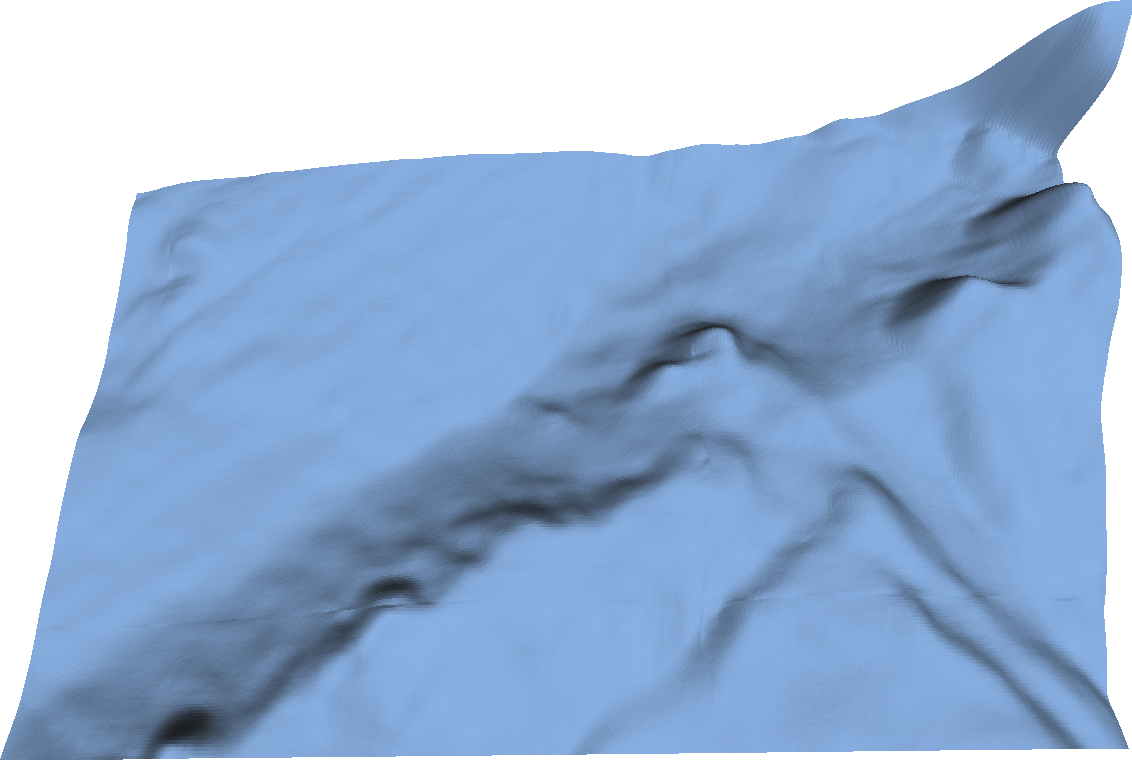}
\end{minipage}%
\begin{minipage}{0.22\columnwidth}
    \centering
    \includegraphics[width=1\columnwidth]{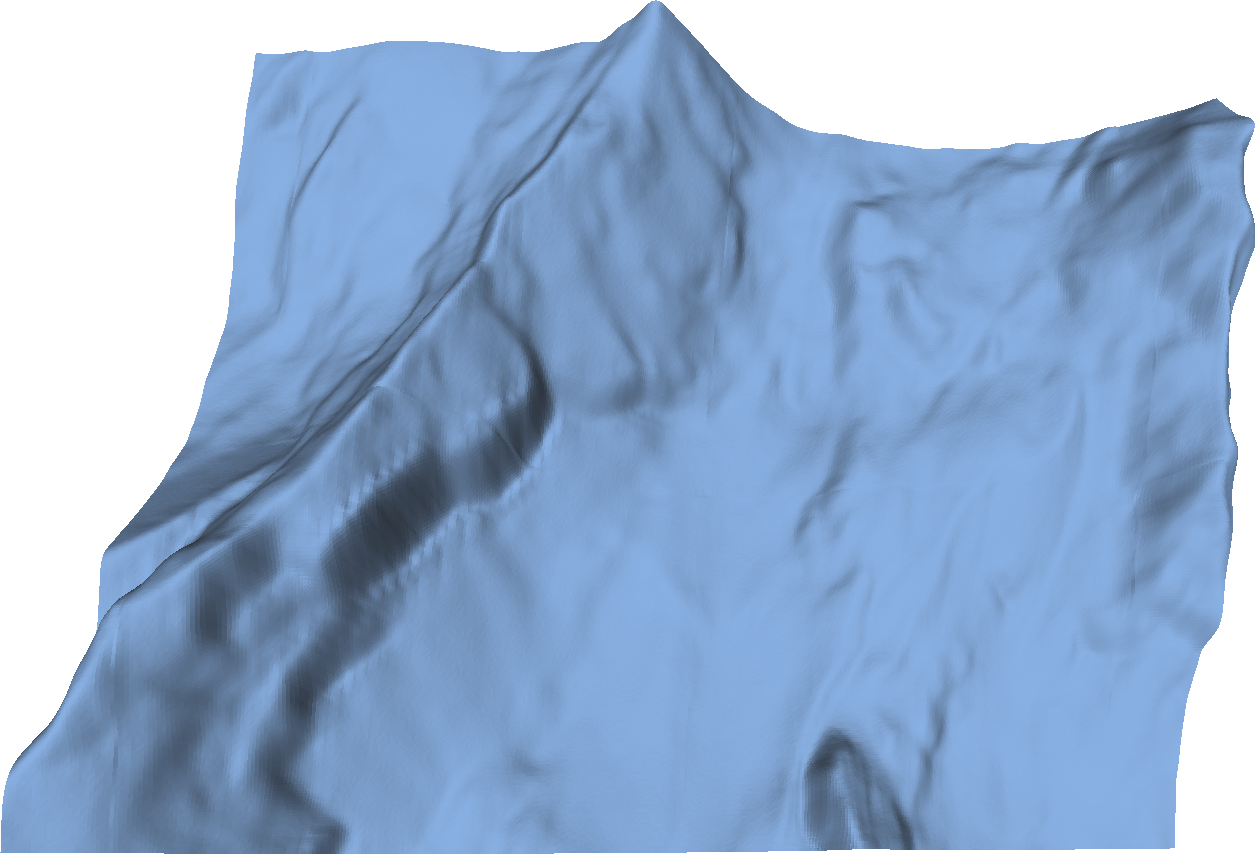}
\end{minipage}%

%for ours atn
\begin{minipage}{0.095\columnwidth}
    \centering
    \text{Ours}
\end{minipage}%
\begin{minipage}{0.26\columnwidth}
    \centering
    \includegraphics[width=1\columnwidth]{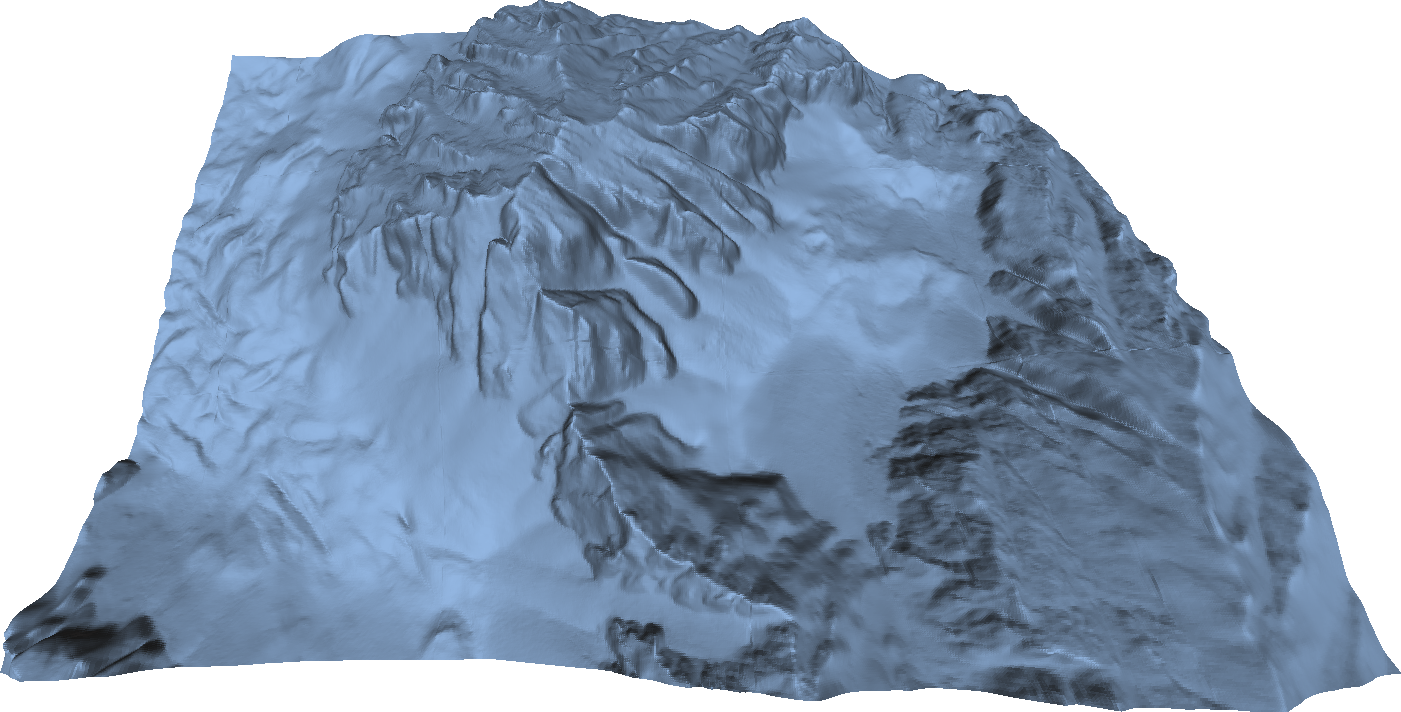} 
\end{minipage}%
\begin{minipage}{0.22\columnwidth}
    \centering
    \includegraphics[width=1\columnwidth]{imgs/forc/forc_atn02.png} 
\end{minipage}%
\begin{minipage}{0.22\columnwidth}
    \centering
    \includegraphics[width=1\columnwidth]{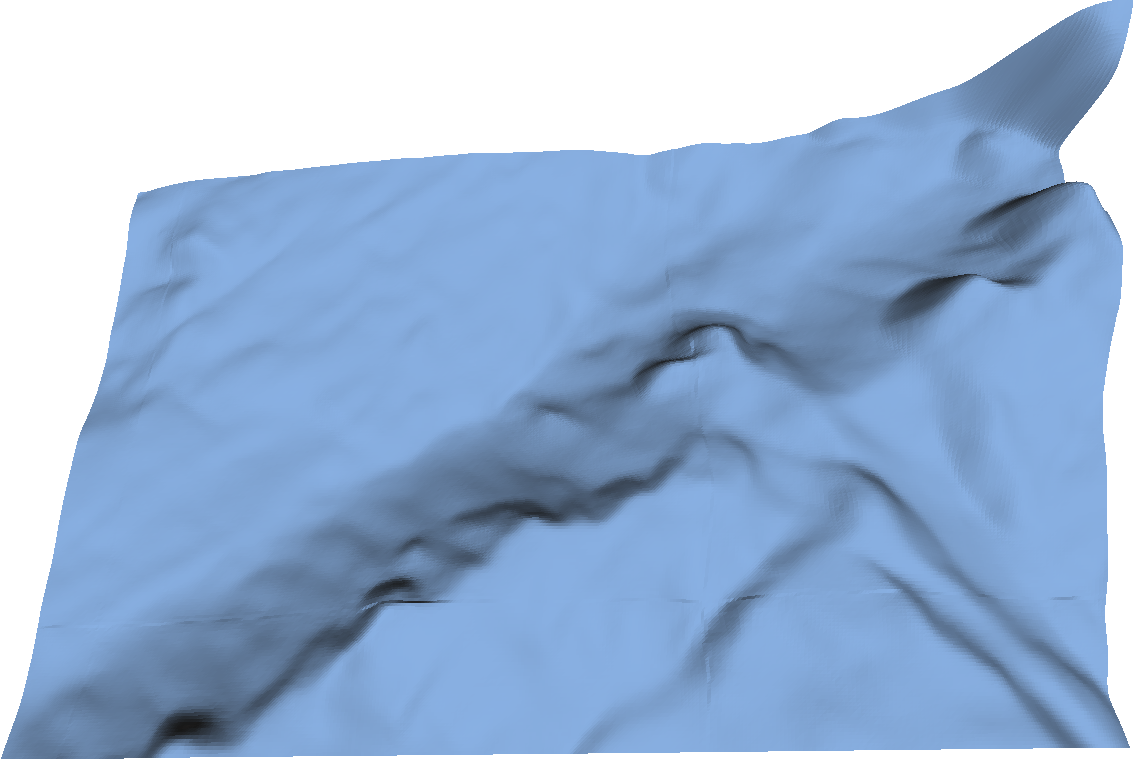} 
\end{minipage}%
\begin{minipage}{0.22\columnwidth}
    \centering
    \includegraphics[width=1\columnwidth]{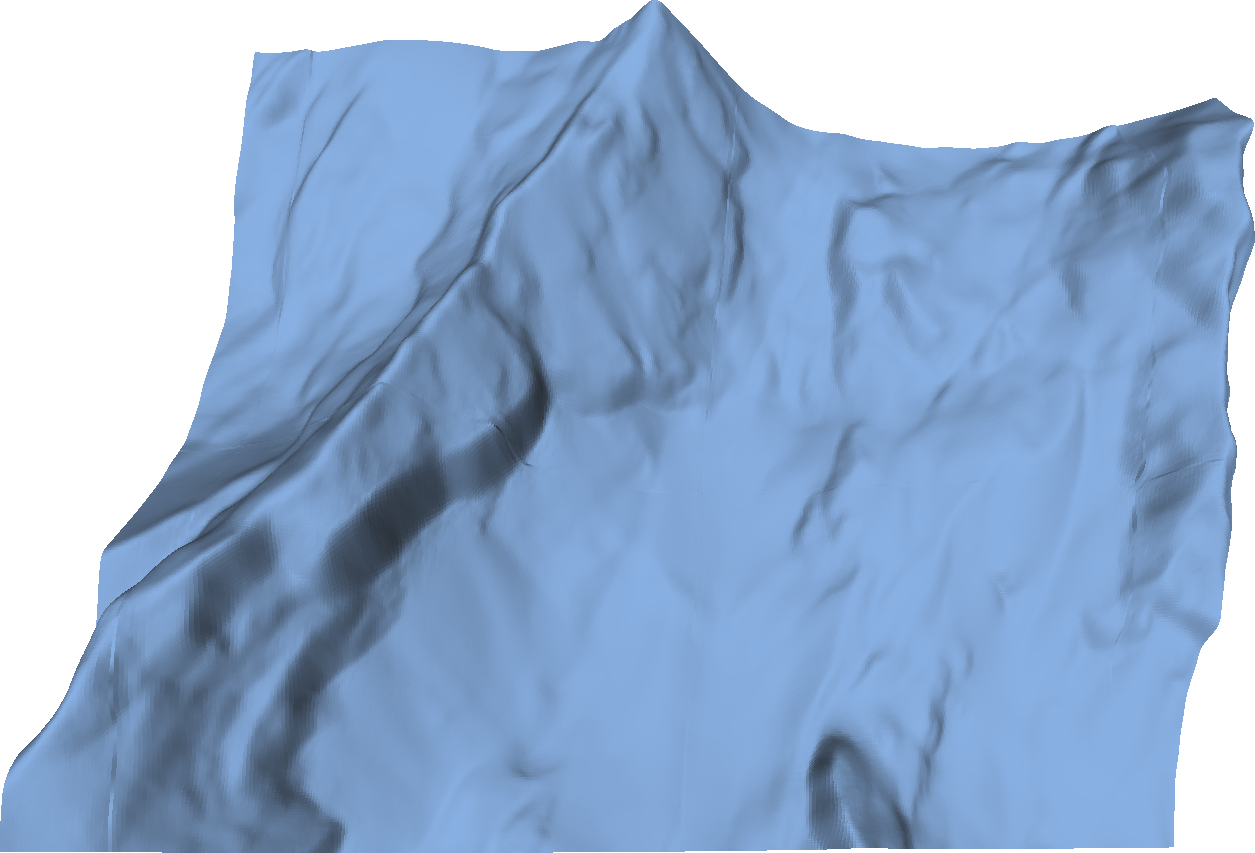} 
\end{minipage}% % atn end

%for ours atn
\begin{minipage}{0.095\columnwidth}
    \centering
    \text{Ground} \\ 
    \text{Truth}
\end{minipage}%
\begin{minipage}{0.26\columnwidth} %GT start
    \centering
    \includegraphics[width=1\columnwidth]{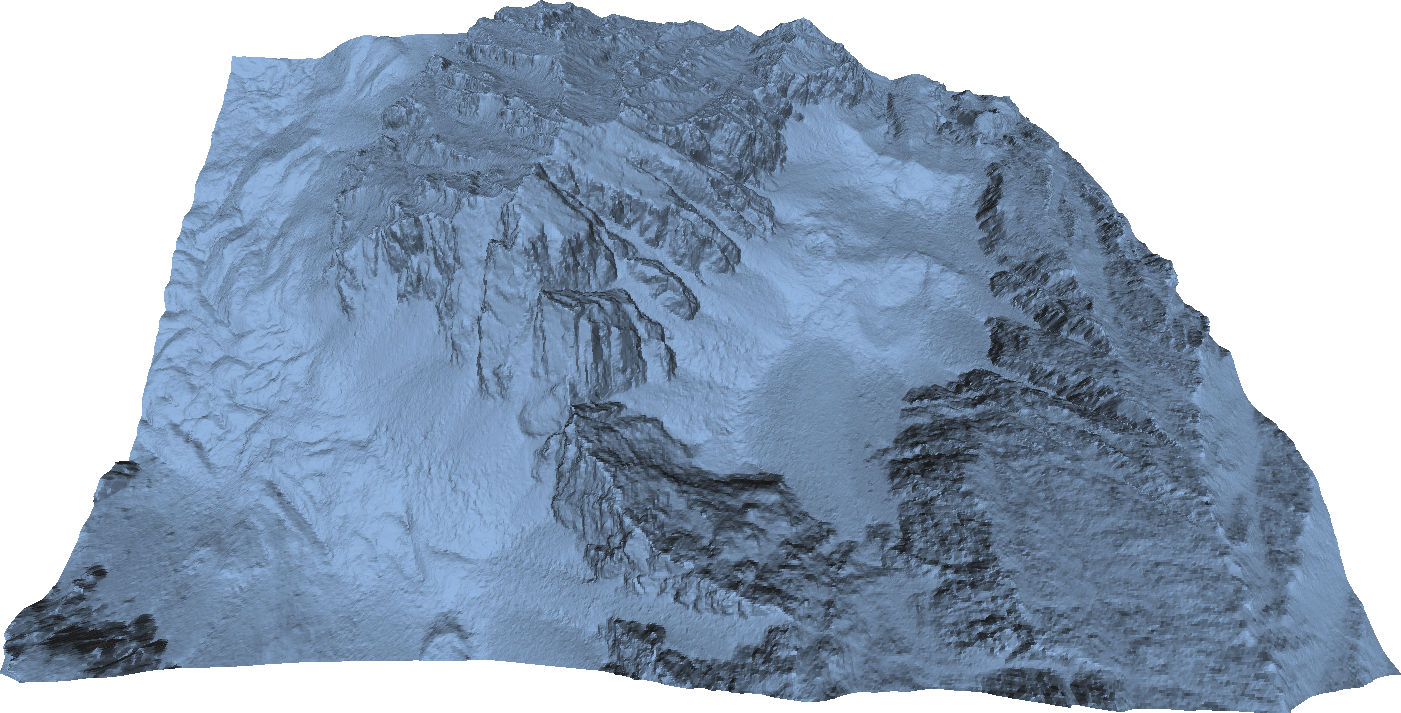} 
\end{minipage}%
\begin{minipage}{0.22\columnwidth}
    \centering
    \includegraphics[width=1\columnwidth]{imgs/forc/forc_gt00.png} 
\end{minipage}%
\begin{minipage}{0.22\columnwidth}
    \centering
    \includegraphics[width=1\columnwidth]{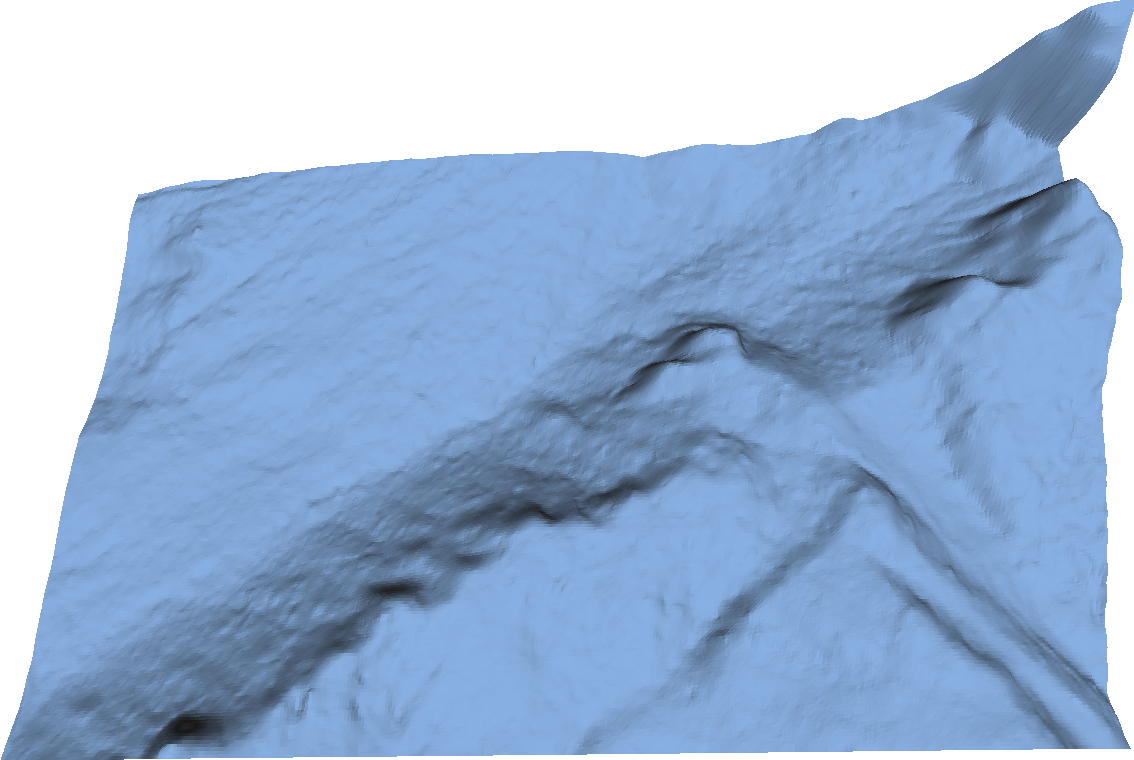} 
\end{minipage}%
\begin{minipage}{0.22\columnwidth}
    \centering
    \includegraphics[width=1\columnwidth]{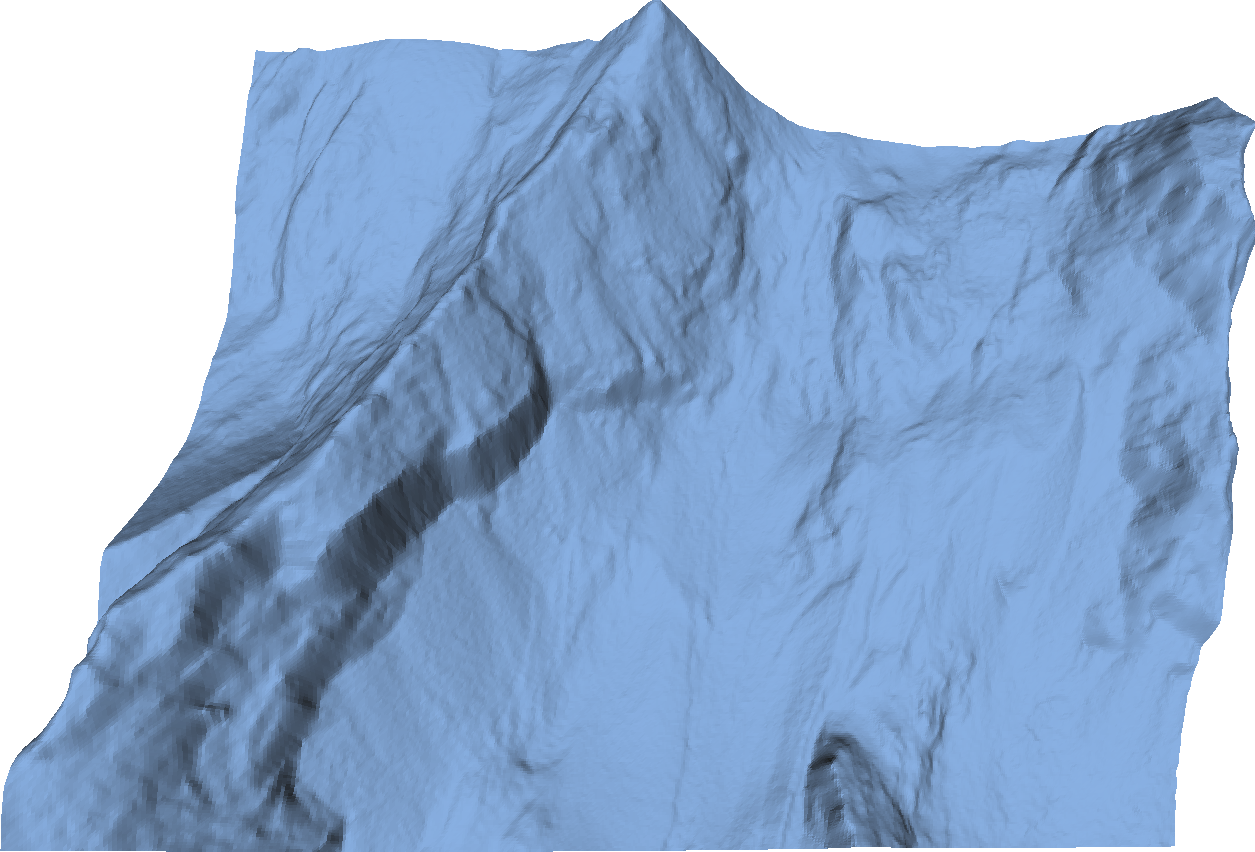} 
\end{minipage}% %GT
\caption{Qualitative comparison of different DEM super-resolution methods}
\label{fig:MeshPlots}
\vspace{-0.20in}
\end{figure}

\begin{table}[h]
\vspace{-0.2in}
\caption{Comparison: PSNR values (The higher the better).}
\centering
\resizebox{\textwidth}{!}{
\begin{tabular}{|c@{\hskip2pt}|c|c|c@{\hskip2pt}|c|c|c|c|}
\hline
Input & \multicolumn{4}{c|}{Only LRDEM} & \multicolumn{3}{c|}{LRDEM and RGB} \\
\hline
Region      & Bicubic   & DSRFB & DSRFO & FCND & FCN & AFN & AFNO \\ \hline
Bassiero    & 60.5 & 62.261   & 62.687 & 62.752 & 63.4 & \textbf{63.958} & \textbf{64.113} \\ \hline
Forcanada   & 58.6 & 60.383 & 60.761 & 60.837 & 62.0 & \textbf{62.351} & \textbf{62.574} \\ \hline
Durrenstein & 59.5 & 63.076 & 63.766 & 63.924  & 63.6 & \textbf{63.841} & \textbf{64.061}\\ \hline
Monte Magro & 67.2 & 70.461 & 71.081 & 71.196  & 71.1 & \textbf{71.211} & \textbf{71.417} \\ \hline
\end{tabular}
}
\label{table:psnr}
\vspace{-0.15in}
\end{table}
%%%%%%%%%%%%%%%%%%%%%%%%%%%%%%%%%%%%%%
%\subsection{Realistic Appearance of Terrains}
 From our test regions, we pick one patch each based on certain geographical property, typically containing one major terrain feature. In Fig. \ref{fig:MeshPlots}, first row shows the aerial view of the selected terrain patches. From Bassiero, we select a highly varying terrain patch. From Forcanada, we choose a patch with bare surface. Patches from Durrenstein and Monte Magro respectively have terrains covered with dense vegetation and snow. From comparison results in Fig. \ref{fig:MeshPlots}, we can see that, for Bassiero, our method is able to recover most of the terrain variations in the terrain. In low resolution input of Forcanada, almost all terrain details have been lost, yet our method can recover most of the lost structure. In cases of covered terrains in Durrenstein and Monte Magro, where LRDEM is seen to have more variations, our method has introduced the least noise. Additional results are available in the supplementary video.
%  \textcolor{red}{The gains achieved by our method (over SOTA) in terms of absolute height values are around 0.5-1.0m which is quite valuable.}
 
%  Please refer to our supplementary material for an extended set of video results.
 
%------------------------------------------------------------------------- 
\subsection{Ablation Studies}
To justify the effectiveness of the Attention module, we thoroughly test our network by creating its variants around Attentional Feedback Module. We discuss four major studies in this section.\\

\noindent
{\bf Without Attention Module: }
 In this experiment, we remove the attention module from the network entirely. For fusing the features from two modalities, i.e. $F_{DEM}$ and $F_{RGB}$, we use channel concatenation followed by $Conv(m,1)$ layer. We keep the rest of the setup same as in AFN. The reduction in performance of the network can be seen in Table \ref{table:ablation} which supports the role of attention module in selective feature extraction.\\

\begin{table}[ht]
    \centering
    \caption{Ablation Studies.}
    \begin{tabular}{|c|c|c|c|c|c|c|c|c|}
        \hline
        \multirow{2}{*}{Region} & \multicolumn{2}{c|}{Without AFM} & \multicolumn{2}{c|}{AFN0} & \multicolumn{2}{c|}{AFN64} & \multicolumn{2}{c|}{AFN} \\
         \cline{2-9}
         & PSNR & RMSE & PSNR & RMSE &PSNR & RMSE &PSNR & RMSE \\
         \hline
        Bassiero & 62.406 & 1.128 & 63.108 & 1.04 & 63.724 & 0.969 & 63.958 & 0.943 \\
        \hline
        Forcanada & 60.537 & 1.303 & 61.355 & 1.186 & 62.141 & 1.084 & 62.351& 1.058 \\
        \hline
        Durrenstein & 62.994 & 0.967 & 63.769 & 0.884 & 64.116 & 0.85 & 63.841 & 0.877 \\
        \hline
        Monte Magro & 70.365 & 0.64 &  70.934 & 0.599 & 71.154 & 0.584 & 71.211 & 0.58 \\
        \hline
    \end{tabular}
    \label{table:ablation}
    \vspace{-0.20in}
\end{table}

\noindent
{\bf Static Attention Masks: }
In AFN, the attention masks for both $F_{RU}$ and $F_{RGB}$ get updated with iterations. In this study, we move the attention module outside the feedback network and use feedback module only for refining the $F_{RU}$ features. So in this case, we denote the attention state as static and call this variant as AFN0. Comparison from Table \ref{table:ablation} confirms that iterative attention can help the network learn more refined feature than fixed attention mask.\\

\noindent
{\bf Number of channels in Attention Module:}
To understand the contribution of AFM in performance gain, we changed hyper-parameters. We reduced the number of channels to 64 throughout the attention module. We denote AFN in this setup as AFN64. The proportional reduction in performance reflects the role of AFM in capturing the higher frequency details. \\

\begin{table}[ht]
\centering
\vspace{-0.25in}
\caption{Performance of AFND i.e. AFN without using aerial images.}
\resizebox{\textwidth}{!}{
\begin{tabular}{|c@{\hskip2pt}|c|c|c|c|c@{\hskip2pt}|c|c|c|}
\hline
\multirow{2}{*}{Region} & \multicolumn{4}{c|}{PSNR (in dB, the higher the better)} & \multicolumn{4}{c|}{RMSE (in meters, the lower the better)} \\ \cline{2-9} 
 & Bicubic & FCND & DSRFB &  AFND & Bicubic & FCND & DSRFB & AFND \\ \hline
Bassiero    & 60.5 & 62.261 & 62.687 & 62.404 &  1.406 & 1.146 & 1.091  & 1.128 \\  \hline
Forcanada   & 58.6 & 60.383 & 60.761 & 60.504 &   1.632 & 1.326 & 1.2702 & 1.308  \\ \hline
Durrenstein & 59.5 & 63.076 & 63.766 & 63.394 &   1.445 & 0.957 & 0.884  & 0.923  \\ \hline
Monte Magro & 67.2 & 70.461 & 71.081 & 70.768 &   0.917 & 0.632 & 0.589  & 0.611 \\ \hline
\end{tabular}
}
\label{table:FCND}
\vspace{-0.20in}
\end{table}

\noindent
{\bf Performance without Aerial Imagery: }
To test the flexibility and limitations of AFN, we study its performance in absence of aerial imagery. Getting aligned pair of aerial image and DEM could be challenging sometimes and hence we analyze the performance of AFN in absence of aerial image. In this exercise, we replace the input aerial image with an uniform prior image of same dimensions. We call this variant as AFND. Table \ref{table:FCND} shows that despite trained with RGB images, while prediction, AFND selectively picks information from DEM modality and perform consistently better than FCND and almost comparable to DSRFB. Of course, DSRFB was designed to work without RGB. The marginal decrease in performance of AFND compared with DSRFB can be attributed partially to the uniform prior acting as noise and causing the attention module to generate a biased attention response.

% \subsection{Fine-tuning VGG layers Vs Freezing VGG Layers}
\section{Conclusion}
We have proposed a  novel  terrain  amplification  method called AFN for generating the DEM super-resolution. It uses  low-resolution DEM and complementary information from corresponding aerial image by computing an attention mask from the attention module along with the feedback network to enhance the performance of the proposed architecture. While this architecture is able to learn well across different terrains, there is a need to further enhance some key features of terrains, especially in regions with high frequency. Hence, there might be a need to explore the use of multi-scale fusion as an extension to the proposed AFN. Also, similar to other computer vision applications, it might be interesting to generate high-resolution DEM using only aerial image as an input.
%\textcolor{red}{In contrast, trying to generate a HRDEM from only RGB is inherently difficult as the height cues are almost absent in land covered with snowfall, vegetation or even shadows by larger structures.}
\bibliographystyle{splncs}
\bibliography{egbib}

%this would normally be the end of your paper, but you may also have an appendix
%within the given limit of number of pages
\end{document}